\pdfoutput=1
\documentclass[fleqn,usenatbib]{mnras}

\usepackage{newtxtext,newtxmath}
\usepackage[T1]{fontenc}

\usepackage[dvipsnames]{xcolor}
\usepackage[normalem]{ulem}
\usepackage{bm}% needed if you want to
\usepackage[utf8]{inputenc}
\usepackage{pgfplots}
\usepackage{tikz}
\usepgfplotslibrary{groupplots}
\pgfplotsset{compat=1.14}
\usepackage{capt-of}
\usepackage{graphicx}% Include figure files

\usepackage{amssymb}

\usepackage{mathrsfs}
\usepackage{amsmath}
\usepackage{hyperref}
\usepackage{lineno}
\usepackage{subfigure}
\usepackage{widetext}

\definecolor{pink}{RGB}{255, 20, 147}      % pink

\usepackage{subfigure}

\newcommand{\Lagr}{\mathcal{L}}
\definecolor{comment}{RGB}{166, 38, 164}

\title[Multi-messenger Emission from Neutron Star Oceans]{Multi-messenger Emission from Tidal Waves in Neutron Star Oceans}

\author[A.G. Sullivan et al.]{
Andrew G. Sullivan,$^{1}$\thanks{E-mail: ags2198@columbia.edu} 
Lucas M. B. Alves,$^{1}$
Georgina O. Spence,$^{2}$
Isabella P. Leite,$^{3}$
\newauthor
Do\u{g}a Veske,$^{1}$
Imre Bartos,$^{4}$
Zsuzsa M\'arka,$^{5}$
Szabolcs M\'arka$^1$
\\
$^{1}$Department of Physics, Columbia University in the City of New York, New York, NY 10027, USA\\
$^{2}$Department of Mathematics, Barnard College of Columbia University in the City of New York, New York, NY 10027, USA\\
$^3$Department of Biomedical Engineering, Columbia University in the City of New York, New York, NY 10027, USA\\
$^{4}$Department of Physics, University of Florida, Gainesville, FL 32611-8440, USA\\
$^{5}$Columbia Astrophysics Laboratory, Columbia University in the City of New York, New York, NY 10027, USA
}

\date{Accepted 2023 January 31. Received 2023 January 9; in original form 2022 June 3}

\pubyear{2023}

\begin{document}
\label{firstpage}
\pagerange{\pageref{firstpage}--\pageref{lastpage}}
\maketitle

\begin{abstract}
Neutron stars in astrophysical binary systems represent exciting sources for multi-messenger astrophysics. A potential source of electromagnetic transients from compact binary systems is the neutron star ocean, the external fluid layer encasing a neutron star. We present a groundwork study into tidal waves in neutron star oceans and their consequences. Specifically, we investigate how oscillation modes in neutron star oceans can be tidally excited during compact binary inspirals and parabolic encounters. We find that neutron star oceans can sustain tidal waves with frequencies between {$0.01-20$} Hz. Our results suggest that tidally resonant neutron star ocean waves may serve as a never-before studied source of precursor electromagnetic emission prior to neutron star-black hole and binary neutron star mergers. If accompanied by electromagnetic flares, tidally resonant neutron star ocean waves, whose energy budget can reach $10^{46}$ erg, may serve as early warning signs ({$\gtrsim 1$ minute} before merger) for compact binary mergers. Similarly, excited ocean tidal waves will coincide with neutron star parabolic encounters. Depending on the neutron star ocean model and a flare emission scenario, tidally resonant ocean flares may be detectable by Fermi and NuSTAR out to $\gtrsim 100$ Mpc with detection rates as high as $\sim 7$ yr$^{-1}$ for binary neutron stars and $\sim0.6$ yr$^{-1}$ for neutron star-black hole binaries. Observations of emission from neutron star ocean tidal waves along with gravitational waves will provide insight into the equation of state at the neutron star surface, the composition of neutron star oceans and crusts, and neutron star geophysics.
\end{abstract}

% -------------------------------------------
\section{Introduction}
\label{sec:Introduction} 
% -------------------------------------------

 With the recent detections of gravitational waves (GWs) from compact binary systems by GW detectors such as LIGO, Virgo, and KAGRA \citep{2015CQGra..32b4001A, 2015CQGra..32g4001L, 2017PhRvL.119p1101A, 2019CQGra..36p5008A, 2021arXiv210801045T, 2021ApJ...915L...5A}, binary systems that include neutron stars have come to the forefront of high energy astrophysics. Neutron stars represent a unique class of stellar objects in that, though very dense, they emit light, making them a candidate for combined GW-electromagnetic multi-messenger astrophysical searches \citep{2015IJMPD..2430012R, 2018LRR....21....3A}. Neutron stars are thought to consist of three distinct layers: a very dense fluid core, a solid crust, and an external fluid ocean \citep{2001ApJ...550..426L}. The respective properties of each of these layers largely depend on the neutron star equation of state \citep{2001ApJ...550..426L}, whose details remain an active problem in nuclear physics and astrophysics. Detections of X-ray bursts \citep{1995ApJ...449..800B, 2014ApJ...793L..38S, 2020MNRAS.491.6032C}, gamma ray bursts, \citep{2012PhRvL.108a1102T, 2013ApJ...777..103T, 2020PhRvD.101h3002S}, ejecta from compact binary inspirals \citep{2010MNRAS.406.2650M, 2014MNRAS.441.3444M, 2017IJMPC..2850080G, 2017LRR....20....3M, 2018ApJ...869..130R, 2019LRR....23....1M, 2017ApJ...848L..16S, 2017ApJ...848L..17C, 2017ApJ...848L..18N, 2017ApJ...848L..19C, 2019Natur.569...85B}, and GWs \citep{1998MNRAS.299.1059A, 2010CQGra..27s4006F, 2018MNRAS.478..167S, 2020GReGr..52..109C} may probe this structure.
 
 In recent years, GW astrophysics has become a unique observational tool to study neutron star physics. The answers to a number of open questions concerning the properties of neutron stars may lie in the rich capabilities of multi-messenger astrophysics with GWs. Works have investigated the possibility of mountains on the surfaces of spinning neutron stars, whose asymmetries could generate detectable continuous GWs \citep{10.1046/j.1365-8711.2000.03938.x,2020MNRAS.494.2839O, 2021MNRAS.500.5570G, 2021MNRAS.507..116G}. Searches for continuous GWs potentially originating from spinning neutron stars have been undertaken \citep{2015ApJ...813...39A, 2020ApJ...897...22P, 2021arXiv211210990T, 2022PhRvD.105h2005A, 2022arXiv220100697T}, and may study neutron star geophysical structure and seismology \citep{2017IJMPC..2850080G, 2018PhRvD..98d4007Y, 2018MNRAS.478..167S, 2021Univ....7...97A}.  
 
Neutron stars exhibit a variety of pulsational modes \citep{1988ApJ...325..725M, 1994ApJ...426..688R, 1994MNRAS.270..611L, 2006PhRvD..73h4010P, 2007CQGra..24.4147S, 2012MNRAS.419..638P}. These oscillation modes are associated with the restoring forces and structure of the star. Modes include the fundamental mode or $f$-mode \citep{2010f-mode,2017PhRvL.119p1101A, 2019PhRvC..99d5806W, 2021ApJ...915L...5A}, pressure modes or $p$-modes \citep{2014MsT..........5B}, gravity modes or $g$-modes \citep{1988ApJ...325..725M,  1995ApJ...449..800B, 1996ApJ...460..827B, 2016ApJ...832...44D, 2020PhRvD.101h3001A, 2021MNRAS.504.1273P, 2021arXiv210700533K}, $r$-modes in rotating neutron stars \citep{2015IJMPE..2441007H, 2015PhDT.......385M, 2021PhRvD.103f3020M, 2020MNRAS.491.6032C}, and interface modes or $i$-modes \citep{1988ApJ...325..725M, 2012MNRAS.419..638P}.  Oscillation modes may be excited during accretion \citep{1994ApJ...426..688R, 2016ApJ...832...44D} or by tides \citep{1994MNRAS.270..611L, 1999MNRAS.308..153H, 2021MNRAS.500.5570G}.
 
Neutron star oscillations have been studied in connection with emission of electromagnetic radiation. The prospect of observing neutron star ocean {oscillations} induced by accretion, in particular, has been considered in many previous works \citep{1995ApJ...449..800B, 1996ApJ...460..827B, 2004ApJ...600..939H, 2016ApJ...832...44D, 2020MNRAS.491.6032C, 2020MNRAS.496.2098V}. Thermonuclear burning on neutron star surfaces during accretion can excite oscillation modes, which could represent the oscillations in type-I X-ray burst light curves 
\citep{1975ApJ...195..735H,1976Natur.263..101W,Maraschi1977,1995ApJ...449..800B, 2002ApJ...566.1018S,2004ApJ...600..914L, 2005ApJ...629..438P, 2018MNRAS.477.4391C, 2020MNRAS.491.6032C}. Observed thermonuclear X-ray bursts on neutron stars show signs of ocean mode oscillation \citep{2008ApJS..179..360G, 2019ApJS..245...19B, 2021MNRAS.508.2123R, 2021ApJ...907...79B}. 

Because neutron stars can exist in binaries, tidal deformations play a role in neutron star physics as well. A neutron star's response to tidal forces largely depends on its internal properties, including its oscillation modes \citep{1994MNRAS.270..611L}. Observations of tidally excited oscillation modes would probe the composition of neutron stars. 

In this work, we analyze neutron star ocean oscillations generated by the dynamical tide during interactions with other compact objects. We principally consider ocean tidal waves in compact binary inspirals, where tidal forces become resonant with neutron star oceans. We also investigate tidal waves from unbound neutron star encounters. We present models for neutron star oceans and investigate the size of tidal waves sustainable in these oceans. Ultimately, we consider astrophysical emission that tidally excited neutron star oceans might produce, including electromagnetic flares and GWs. We perform all of our analysis using Newtonian theory due to the exploratory and phenomenological nature of this study. 
 
 We divide the paper into the following sections. In section \ref{sec:backgroundNS}, we present the background neutron star model used, as well as introduce the three neutron star ocean models investigated. In section \ref{sec:OscModes}, we discuss the equations of motion for neutron star oscillations and determine the neutron star ocean oscillation modes for our models. In section \ref{sec:Tides}, we discuss the tidal interaction and compute tidal wave properties for each of the oceans and orbital configurations considered. In section \ref{sec:discussion}, we discuss our results and their consequences, including potential emission produced by neutron star ocean tidal waves. In section \ref{sec:conclusion}, we conclude.
% -------------------------------------------

% -------------------------------------------
\section{Background Neutron Star and Ocean Model}
\label{sec:backgroundNS}
To focus on the properties of the neutron star ocean, we use a simple background neutron star model with a rigid crust. {We will later extrapolate our results with this model to the case where the neutron star crust is elastic rather than rigid.}

To solve for the star's background density $\rho$ and pressure $p$, we use the classical equilibrium equations for a spherically symmetric fluid
\begin{subequations}
\label{eq:backgroundfluid}
\begin{equation}
\label{eq:backgroundfluid1}
    \frac{dp}{dr}=-\rho g,
\end{equation}
\begin{equation}
    \frac{dM}{dr}=4\pi G r^2 \rho,
\end{equation}
\end{subequations}
where $M(r)$ is the mass enclosed at a given radius, $G$ is Newton's gravitational constant, and $g=\frac{G M(r)}{r^2}$ \citep{chandrasekhar1957introduction}. Given an equation of state, these equations can be solved and provide the star's background pressure and density. In this work, we use a polytropic equation of state \citep{2010IJMPD..19.1569F}
\begin{equation}
\label{eq:EOS}
    p=K \rho^\Gamma,
\end{equation}
where $K$ is a proportionality constant. Choosing $\Gamma=2$ yields an analytic solution for the mass density when $r< R_{\star}$
\begin{equation}
\label{eq:densityrad}
    \rho(r)=\rho_c \frac{\sin{\sqrt{\frac{4\pi G}{2 K}} r}}{\sqrt{\frac{4\pi G}{2 K}} r},
\end{equation}
where $\rho_c$ is the density at the center of the star and $R_{\star}$ is the radius of the neutron star  \citep{chandrasekhar1957introduction}. When $r> R_{\star}$, we have $\rho(r)=0$. Note that the radius of the star is completely specified by the constant $K$.

For this study, we assume our neutron star is non-rotating and has no magnetic field. The effects of rotation and magnetization, if small enough, will serve as perturbations to the oscillation mode structures and frequencies without changing the physics  \citep{2021MNRAS.506.2985K, 2021MNRAS.508.1732K, 2021FrASS...8..166K}. Because we are interested in early inspirals, the effects of general relativity should not play a role in spinning up the rotation of neutron stars.  While we expect effects such as tidal locking to also spin up neutron stars, we do not consider them in our study. We leave consideration of rotating and magnetized neutron stars to future work.

\subsection{Neutron Star Ocean Depth}
The depth of the neutron star ocean depends on the density at which the neutron star crust melts. The top of the crust is typically considered to be a body-centered cubic (bcc) Coulomb crystal \citep{1995ApJ...449..800B, 2007ASSL..326.....H,
2009PhRvL.102s1102H, 2018MNRAS.480.5511B, 2020PhRvD.101j3025G}. In a Coulomb crystal, the ions which compose the lattice interact exclusively by the Coulomb interaction \citep{2018MNRAS.477.4391C} because the electron screening in the outer crust is weak \citep{2008LRR....11...10C}. The Coulomb crystal undergoes a phase transition when the thermal energy exceeds the electric binding energy of the material by some critical factor $\gamma$ \citep{1993PhRvE..47.4330F}. The crust melts when the following condition is met:
\begin{equation}
\label{eq:melt}
    k_{B}T\geq\frac{1}{\gamma} \frac{1}{4 \pi \epsilon_0}\frac{Z^2e^2}{d},
\end{equation}
where $k_{B}$ is Boltzmann's constant, $T$ is the temperature, $\epsilon_0$ is the permittivity of free space, Z is the proton number of atomic nuclei in the lattice, $e$ is electron charge, and $d$ is the mean spacing between nuclei. Molecular dynamics studies have found $\gamma\approx173$ \citep{1993PhRvE..47.4330F}.  Assuming that the ion number density is $n_i=(\frac{4}{3}\pi d^3)^{-1}$, the mass density at which the crust melts and the ocean forms is
\begin{equation}
\begin{split}
\label{eq:odensity}
    \rho_{o}=&A m_n n_i=\frac{3}{4 \pi} A m_n \left(\gamma\frac{4\pi\epsilon_0k_B T}{Z^2 e^2}\right)^3 \\& \approx2.705\times10^{10}\text{ g cm}^{-3} \left(\frac{A}{16}\right) \left(\frac{8}{Z}\right)^6 \left(\frac{T}{10^8 K}\right)^3, 
\end{split}
\end{equation}
where $A$ is the atomic mass of the nuclei in the lattice, $m_n$ is nucleon mass, and we have used the condition in equation \ref{eq:melt} for $d$ at the transition between the neutron star crust and ocean. Equation \ref{eq:odensity} shows the melting density's strong dependence on temperature and ion atomic number. More proton-rich nuclei will reduce the density at which the ocean begins. 

By plugging equation \ref{eq:odensity} into the left-hand side of equation \ref{eq:densityrad}, we determine the radius at which the ocean begins and by extension the depth of the ocean as a function of $A$, $Z$, and $T$ when $\Gamma=2$. Since the ocean is very shallow compared to the neutron star radius \citep{1995ApJ...449..800B, 1996ApJ...460..827B, 2004A&A...421L...5U, 2016ApJ...832...44D,2020MNRAS.496.2098V}, we approximate $r$ in the denominator of equation \ref{eq:densityrad} as the stellar radius {$R_{\star}=\sqrt{\frac{2 K}{4 \pi G}}\frac{\pi}{2}$}. The radius at which the neutron star ocean begins for a $\Gamma=2$ polytropic equation of state is
\begin{equation}
    r_o=\sqrt{\frac{2K}{4\pi G}}\left[\arccos{\left(\frac{3}{8} \frac{A m_n}{\rho_c} \left(\gamma\frac{4\pi\epsilon_0k_B T}{Z^2 e^2}\right)^3\right)}+\frac{\pi}{2}\right].
    \label{eq:oceandepth}
\end{equation}
{We also obtain an approximate ocean depth $h_o$ for a general polytropic equation of state in terms of $\rho_o$. Differentiating equation \ref{eq:EOS} gives
\begin{equation}
\label{eq:diffeos}
    \frac{dp}{dr}=\Gamma K \rho^{\Gamma-1}\frac{d \rho}{d r}.
\end{equation}
Combining equations \ref{eq:backgroundfluid1} and \ref{eq:diffeos} provides a differential equation for $\rho$ and $r$
\begin{equation}
\label{eq:drhodh}
    \frac{d \rho}{dr}=-\frac{g}{\Gamma K}\rho^{2-\Gamma}.
\end{equation}
Integrating equation \ref{eq:drhodh} from the ocean floor to the surface assuming constant $g=\frac{G M}{R_\star^2}$ gives
\begin{equation}
\label{eq:oceandepgen}
    h_o=R_\star-r_o=\frac{\Gamma K}{\Gamma-1}\frac{\rho^{\Gamma-1}_o}{g}.
\end{equation}
Any choice of $K$ and $\Gamma$ in the ocean can therefore give an approximate $h_o$.}

In this work, we consider three model crusts respectively made up of three elements thought to be found in neutron star surfaces due to their production by r-processes \citep{2018JPhG...45i3001M}: carbon with Z = 6 and A = 12, oxygen with Z = 8 and A = 16, and iron with Z = 26 and A = 56. For referential convenience, we refer to the three oceans corresponding to these differently composed crusts as carbon, oxygen, or iron oceans. Neutron star crust temperatures are typically $T\sim10^7$ K when the crust is in thermal equilibrium with the core \citep{1998ApJ...504L..95B, 2009ApJ...698.1020B}. Accretion can raise the temperature of the neutron star ocean floor to $T\sim10^8$ K \citep{1984ApJ...278..813F, 1990A&A...227..431H, 2003A&A...404L..33H, 2008A&A...480..459H}. The temperature decreases through the ocean to $10^6$ K at the surface \citep{1990ApJ...362..572M, 2008LRR....11...10C}. In our study, we neglect effects of the ocean temperature gradient. We discuss our choice for crust temperature in section \ref{subsubsec:parameters}.

\section{Neutron Star Ocean Oscillation Modes}
\label{sec:OscModes}
We solve for the dynamical response of the neutron star ocean. To do this, we use the formalism of Lagrangian perturbation theory for fluids \citep{1978ApJ...221..937F}. Using the Newtonian formalism typically used, we solve for the oscillation modes of the neutron star ocean \citep{1971AcA....21..289D, 1974IAUS...59..135L, 1988ApJ...325..725M,2012MNRAS.419..638P}, so that we may study the dynamical response to tidal forces \citep{1994MNRAS.270..611L, 1994ApJ...426..688R, 2012PhRvL.108a1102T, 2013ApJ...777..103T, 2021MNRAS.504.1273P}. 
\subsection{Equations of Motion}
\label{subsec:ModeEOM}
The equation of motion for Lagrangian perturbative displacements is the perturbed Euler equation
\begin{equation}
\label{eq:fluid}
    \partial_t^2 \vec{\xi} + \frac{\nabla \delta p}{\rho}-\frac{\delta \rho}{\rho^2}\nabla p+\nabla\delta\phi-\frac{1}{\rho} \nabla \cdot \boldsymbol{\sigma}=-\nabla \chi ,
\end{equation}
where $\vec{\xi}$ is the Lagrangian displacement vector, $\rho$ is the background fluid density, $p$ is the background fluid pressure, $\delta \rho$ is the Eulerian perturbation of the density, $\delta p$ is the Eulerian perturbation of the pressure, $\delta \phi$ is the Eulerian perturbation of the gravitational potential, {$\boldsymbol{\sigma}=\sigma_{ij}$ is the elastic stress tensor,} and $\chi$ is an unspecified (for now) external potential that drives the system. {The elastic stress tensor is defined as
\begin{equation}
   \sigma_{ij}=\breve{\mu}\left(\nabla_i \xi_j+\nabla_j\xi_i\right)-\frac{2}{3}\breve{\mu}\delta_{ij}\left(\nabla \cdot \vec{\xi}\right),
\end{equation}
where $\breve{\mu}$ is the shear modulus and $\delta_{ij}$ is the Kronecker delta. In a fluid, $\breve{\mu}=0$.}

The Lagrangian perturbation for density can be written as 
\begin{equation}
\label{eq:density2}
    \Delta \rho=\delta \rho + \vec{\xi}\cdot \nabla \rho =-\rho \nabla \cdot \vec{\xi},
\end{equation}
where the first equality is the definition of the Lagrangian perturbation in terms of the Eulerian perturbation and the second equality arises from conservation of mass \citep{1978ApJ...221..937F}. If the oscillations are adiabatic, the Lagrangian perturbations for pressure and density are related by
\begin{equation}
\label{eq:adiabetic}
    \frac{\Delta \rho}{\rho}=\frac{1}{\Gamma_1}\frac{\Delta p}{p},
\end{equation}
where $\Gamma_1$ is the adiabatic index. We note that $\Gamma_1$ does not necessarily equal $\Gamma$. When $\Gamma_1\neq\Gamma$, the neutron star is stratified and can sustain internal $g$-modes \citep{1988ApJ...325..725M, 1995ApJ...449..800B, 1996ApJ...460..827B, 2020PhRvD.101h3001A, 2021MNRAS.504.1273P}. 

The final equation of motion that governs this system is the perturbative form of the Poisson equation
\begin{equation}
    \nabla^2 \delta \phi= 4\pi G \delta\rho.
\end{equation}
Since the ocean is the uppermost layer of the star and typically very shallow, the perturbation of the gravitational potential $\delta \phi$ and its gradient $\nabla \delta \phi$ must be very small compared to the background gravitational potential $\phi$ and the background gravitational acceleration $g$. As such, we employ the Cowling approximation \citep{1941MNRAS.101..367C}, which approximates $\delta \phi\approx0$ and $\nabla \delta \phi\approx0$. Consequently, we neglect the appearance of $\delta \phi$ in equation \ref{eq:fluid}.

In this section, we set $\chi=0$ and study the homogeneous solutions to equation \ref{eq:fluid}. First, we define a perturbation to the chemical potential per nucleon mass $\delta \Tilde{\mu}$ which, {in a barotropic fluid where $\Gamma=\Gamma_1$}, is related to the perturbation of pressure by
\begin{equation}
\label{eq:deltap}
    \delta p= n_N \delta \mu =\rho \delta \Tilde{\mu},
\end{equation}
where $\delta \mu$ is the normal chemical potential, {and $n_{N}$ is the nucleon number density}. Inserting equation \ref{eq:deltap} for $\delta p$ into equation \ref{eq:fluid}, {explicitly writing out the derivative of the stress tensor}, and applying the Cowling approximation gives
\begin{equation}
\begin{split}
   \label{eq:fluid2}
    0&=\partial_t^2 \vec{\xi} + \nabla \delta \Tilde{\mu}\\&-\frac{1}{\rho}\left(\frac{d \breve{\mu}}{dr}(\nabla_r \vec{\xi}+\nabla \xi_r)+\breve{\mu}\left(\nabla \left(\nabla \cdot \vec{\xi}\right)+ \nabla^2 \vec{\xi}\right)-\frac{2}{3}\nabla\left(\breve{\mu}\nabla \cdot \vec{\xi}\right)\right).
    \end{split}
\end{equation}
We use equations \ref{eq:density2} and \ref{eq:adiabetic} to obtain the other equation we need to solve this system. From the definition of the Lagrangian perturbation \citep{1978ApJ...221..937F}, we have 
\begin{equation}
\label{eq:deltapnew}
\Delta p= \delta p + \vec{\xi} \cdot \nabla p =\rho \delta \Tilde{\mu} + \xi_r  \frac{dp}{dr},
\end{equation}
where $\xi_r$ is the radial component of $\vec{\xi}$ and the second equality comes from equation \ref{eq:deltap} and spherical symmetry (i.e. $\nabla p =\frac{dp}{dr}$). Substituting equations \ref{eq:deltapnew} and \ref{eq:density2} into equation \ref{eq:adiabetic}, we obtain
\begin{equation}
\label{eq:gradxi}
    \nabla \cdot \vec{\xi}= -\frac{1}{\Gamma_1 p}\left(\rho \delta \Tilde{\mu} + \xi_r  \frac{dp}{dr}\right).
\end{equation}
Equations \ref{eq:fluid2}
and \ref{eq:gradxi} are a system of partial differential equations which can be solved using a clever ansatz for $\vec{\xi}$. We decompose $\vec{\xi}$ into normal modes \citep{1971AcA....21..289D,1974IAUS...59..135L, 1988ApJ...325..725M, 2012MNRAS.419..638P}
\begin{equation}
    \vec{\xi}=\sum_{n} e^{i \omega_{n} t} \vec{\xi}_{n} ,
\end{equation}
where $\omega_{n}$ is the angular frequency of a resonant mode and $\vec{\xi}_n$ is the eigenfunction that solves the equation
\begin{equation}
\label{eq:operator}
    (\Lagr-\rho \omega^2)\vec{\xi}=0,
\end{equation}
where $\Lagr$ is an operator defined such that $\Lagr\vec{\xi}=\rho\nabla\delta \Tilde{\mu}-\nabla \cdot \boldsymbol{\sigma}$ \citep{1977ApJ...213..183P, 2021MNRAS.504.1273P}.
The index $n$ denotes the mode. The orthogonality of these modes requires \citep{1977ApJ...213..183P, 1994MNRAS.270..611L, 2021MNRAS.504.1273P}
\begin{equation}
\label{eq:norm}
    \langle \vec{\xi}_n | \vec{\xi}_m \rangle = \int \rho \vec{\xi}^*_n \cdot \vec{\xi}_m dV=A^2_{n}\delta_{nm},
\end{equation}
where the integral is over the volume of the star, $\vec{\xi}^*_n$ is the complex conjugate of $\vec{\xi}_n$, and $A^2_{n}$ is the normalization factor.
{The spherical symmetry of the problem allows us to write} $\vec{{\xi}_n}$ as \citep{1971AcA....21..289D,1974IAUS...59..135L, 1988ApJ...325..725M, 2012MNRAS.419..638P}
\begin{equation}
\label{eq:guess}
   \vec{\xi}_n=\left(U(r) Y_{l m}(\theta, \phi), V(r) \partial_\theta Y_{l m}(\theta, \phi),  \frac{V(r)}{\sin{\theta}}\partial_\phi Y_{l m}(\theta, \phi)\right),
\end{equation}
where $Y_{l m}(\theta, \phi)$ are the spherical harmonic functions \citep{1962clel.book.....J}, $\partial_x$ is the partial derivative with respect to the variable $x$, and $U(r)$ and $V(r)$ are functions of the radial coordinate that must be solved for. {The spherical symmetry also allows us to write} $\delta \Tilde{\mu}$ as $\delta \Tilde{\mu}=\delta \Tilde{\mu}(r) Y_{l m}(\theta, \phi) e^{i \omega t}$.

\subsubsection{Fluid Ocean}

{In the fluid components of the neutron star where $\breve{\mu}=0$, equation \ref{eq:fluid2} simplifies considerably.} With spherical symmetry and $\breve{\mu}=0$, equations \ref{eq:fluid2} and \ref{eq:gradxi} become first order ordinary differential equations in the radial coordinate $r$
\begin{subequations}
\label{eq:PDEs}
\begin{equation}
    -\omega^2 U+\frac{d}{dr}{\delta \Tilde{\mu}}=0,
    \label{eq:Udmr}
\end{equation}
\begin{equation}
\label{eq:Vchem}
    -\omega^2 V+ \frac{\delta \Tilde{\mu}}{r}=0,
\end{equation}
\begin{equation}
   \frac{dU}{dr}+\frac{2}{r}U-\frac{l(l+1)}{r}V=-\frac{1}{\Gamma_1 p}\left(\rho\delta \Tilde{\mu} + U  \frac{dp}{dr}\right).
   \label{eq:dUr}
\end{equation}
{These equations are valid in the fluid neutron star ocean.}

\end{subequations}
Rearranging equations \ref{eq:Udmr} and \ref{eq:dUr} and using the relationship in equation \ref{eq:Vchem}, this system reduces to two ordinary differential equations  
\begin{subequations}
\begin{equation}
    \frac{d}{dr}{\delta \Tilde{\mu}}=\omega^2 U,
\end{equation}
\begin{equation}
   \frac{dU}{dr}= -\left(\frac{2}{r}+\frac{1}{\Gamma_1 p}\frac{dp}{dr}\right) U+\left(-\frac{\rho}{\Gamma_1 p}+\frac{l(l+1)}{r^2 \omega^2}\right)\delta \Tilde{\mu}.
\end{equation}
\end{subequations}
We write these equations in terms of the dimensionless variables $y_1=\frac{U}{r}$ and $y_2=\frac{\delta \Tilde{\mu}}{r g}$, where $g$ is the background gravitational field as a function of radius. We obtain 
\begin{subequations}
\label{eq:ODEs}
\begin{equation}
   \frac{dy_1}{dr}= -\left(\frac{3}{r}+\frac{1}{\Gamma_1 p}\frac{dp}{dr}\right) y_1+\left(-\frac{\rho}{\Gamma_1 p}+\frac{l(l+1)}{r^2 \omega^2}\right)g y_2,
\end{equation}
\begin{equation}
    \frac{dy_2}{dr}=\frac{\omega^2}{g} y_1-\left(\frac{1}{r}+\frac{1}{g}\frac{dg}{dr}\right)y_2.
\end{equation}
\end{subequations}
In the single fluid limit and the Cowling approximation, the equations in section A1 of \cite{2012MNRAS.419..638P} reduce to equations \ref{eq:ODEs}. Given boundary conditions at the ocean-crust interface and ocean surface, a value for $\Gamma_1$ (which is not constant in general), and a value of $l$, we may solve these equations as a boundary value problem. 

\subsubsection{Elastic Crust}
{If the crust is elastic, the ocean oscillation modes may penetrate into the crust, requiring one to solve equations \ref{eq:fluid2} when $\breve{\mu}\neq0$ \citep{2005ApJ...619.1054P}. In this work, rather than solving equations \ref{eq:fluid2} and \ref{eq:gradxi} in the elastic crust, we will solve for modes in the ocean assuming a rigid crust and extrapolate our results from the rigid to the elastic case, focusing particularly on the consequences for the neutron star tide.}

\subsection{Boundary Conditions}
{In this work, we solve equations \ref{eq:PDEs} assuming the mode is entirely confined to the ocean.
For this simple case, oscillations do not penetrate into the crust.} At the ocean floor we apply the condition
\begin{equation}
\label{eq:rigidcrust}
    y_1(r_o)=0.
\end{equation}
This is the same condition applied by \cite{1995ApJ...449..800B} to solve for deep ocean $g$-modes. At the surface of the ocean we apply the condition $\Delta p=0$ \citep{1988ApJ...325..725M, 1994MNRAS.270..611L}. In our variables, this becomes 
\begin{equation}
    0=y_1(R_{\star})-y_2(R_{\star}).
    \label{eq:surfacerel}
\end{equation}
Additionally, so that we can find the functional form of the oscillation modes, we apply a normalization condition at the surface of the ocean, 
\begin{equation}
    y_1(R_{\star})=1.
    \label{eq:surfacenorm}
\end{equation}
With these three boundary conditions, our system is closed and solvable. {We note that equation \ref{eq:rigidcrust} only preserves continuity of the radial displacement if there is no radial displacement in the crust. This is not necessarily a suitable boundary condition for an elastic crust as the true ocean-crust junction condition is the continuity of the radial displacement and traction variables \citep{ 1988ApJ...325..725M, 2012MNRAS.419..638P, 2021MNRAS.504.1273P}. For a more detailed treatment including the mode's penetration into the crust, one must impose these conditions.}

\subsection{Semi-Analytic Ocean Modes and Tidal Resonance}
We now provide a simple analytic argument {to demonstrate the existence of ocean modes and estimate} how the ocean mode frequency scales with model parameters {in both the rigid crust and elastic crust cases}. This will also give the time of tidal resonance as a function of model parameters.
\subsubsection{Shallow Ocean Surface Wave Model}
{Treating the neutron star ocean as an incompressible shallow ocean, we analytically estimate the neutron star ocean mode frequencies. For waves in a shallow ocean, one solves for the height of the wave above the ocean surface $\eta$. In an incompressible fluid, the density $\rho$ is not a function of the pressure, so the pressure is often taken to be $p=\rho g(h_f-z)$, where $h_f$ is the total height of the ocean. When the height is perturbed by surface waves, we have $h_f=h_o+\eta$, where $h_o$ is the equilibrium depth of the ocean. For waves with $\eta<<h_o$, the perturbed Euler equation and the continuity equation become
 \begin{subequations}
   \begin{equation}
   \label{eq:eulershallow}
       \partial_t \vec{v}_H=-\frac{\nabla_H \delta p}{\rho},
   \end{equation}
   \begin{equation}
   \label{eq:continuityshallow}
       \rho \partial_t \eta+\rho h_o \nabla_H \cdot \vec{v}_H=0,
   \end{equation}
 \end{subequations}
 where $\vec{v}_H=(v_x, v_y)$ is the fluid velocity in the horizontal direction, the gradient $\nabla_H=(\partial_x, \partial_y)$ is the gradient in the horizontal direction, and $\delta p$ is the perturbation to the pressure due to the wave \citep{Randall2006TheSW}.}
 
Because $h_f=h_o+\eta$ for a perturbed ocean, we have $p+\delta p=\rho g(h_o+\eta-z$). $p=\rho g(h_o-z)$ is the background pressure, so the perturbation to the pressure is $\delta p=\rho g \eta$. Equation \ref{eq:eulershallow} becomes
\begin{equation}
\label{eq:eulershalloweta}
     \partial_t \vec{v}_H=-g \nabla_H \eta.
\end{equation}
Taking the divergence of equation \ref{eq:eulershalloweta} gives
\begin{equation}
    \label{eq:diveulershallow}
    \partial_t (\nabla_H\cdot \vec{v}_H)=-g \nabla^2_H \eta.
\end{equation}
Dividing out the mass density and taking the time derivative of equation \ref{eq:continuityshallow} gives
\begin{equation}
   \label{eq:continuityshallow2}
     \partial^2_t \eta+ h_o \partial_t(\nabla_H \cdot \vec{v}_H)=0.
   \end{equation}
Combining equations \ref{eq:diveulershallow} and \ref{eq:continuityshallow2} gives the wave equation for the height of the wave $\eta$
\begin{equation}
\label{eq:waveshallow}
    \partial^2_t \eta-gh_o \nabla^2_H \eta=0.
\end{equation}
At this point, we reintroduce the spherical nature of this problem by assuming $\eta=\eta(t)Y_{l m}$. In spherical coordinates, this problem becomes that of an incompressible fluid shell surrounding a sphere of radius $R_\star,$ similar to a neutron star ocean. While we have previously been working in Cartesian coordinates, the wave equation for $\eta$ holds in spherical coordinates if $\eta$ is not a function of $r$. Expanding $\eta$ in spherical harmonics, equation \ref{eq:waveshallow} becomes
\begin{equation}
\label{eq:waveshallowf}
    \partial^2_t \eta+\frac{l(l+1)}{R_\star^2}gh_o  \eta=0.
\end{equation}
We arrive at ocean mode frequencies in an incompressible fluid ocean surrounding a spherical body of radius $R_\star$ as \begin{equation}
\label{eq:incomprressibleo}
    \omega_i=\frac{1}{R_\star}\sqrt{l(l+1)gh_o}={\sqrt{l(l+1)\frac{G M_\star}{R_\star^3}\frac{h_o}{R_\star}}},
\end{equation}
where the $i$ subscript refers to incompressibility.
To obtain intuition about the functional dependence of the real ocean mode frequencies on our model parameters, we substitute equation \ref{eq:oceandepgen} for $h_o$ in equation \ref{eq:incomprressibleo} and obtain 
\begin{equation}
    \omega\sim\frac{1}{R_\star}\sqrt{l(l+1)\frac{K\Gamma}{\Gamma-1} \rho_o^{\Gamma-1}},
\end{equation}
where we have replaced the equal sign with a tilde for more realistic neutron star oceans.
We now substitute in the expression for $\rho_o$ from equation \ref{eq:odensity} to obtain
\begin{equation}
   \label{eq:estimatesigma}
    \omega\sim\frac{1}{R_{\star}} \sqrt{\frac{l(l+1)K \Gamma}{\Gamma-1}} \left(\gamma\left(\frac{3}{4\pi}m_n\right)^{\frac 13}\frac{4\pi\epsilon_0k_B }{ e^2}\right)^{\frac{3\Gamma-3}{2}} A^{\frac{\Gamma-1}{2}} Z^{3-3\Gamma} T^{\frac{3\Gamma-3}{2}}.
\end{equation}
{Equation \ref{eq:estimatesigma} estimates the mode frequency when the crust is taken to be rigid.}
One can see that the mode frequency increases as a function of $T$ and $A$ and decreases as a function of $Z$ for $\Gamma>1$.

{\cite{2005ApJ...619.1054P} showed that if the pressure at the crust-ocean interface $p_o$ exceeds the shear modulus $\breve{\mu}$, one cannot treat this mode as purely a shallow ocean surface wave, but rather as an interface mode with a nonzero $\vec{\xi}$ in the crust. The interface mode frequency will be the shallow ocean mode frequency scaled by $\sqrt{\frac{\breve{\mu}}{p_o}}$
\begin{equation}
    \label{eq:estimatesigmacrust}
    \omega\sim\left(\frac{\breve{\mu}}{p_o}\right)^\frac{1}{2}\frac{1}{R_{\star}} \sqrt{\frac{l(l+1)K \Gamma}{\Gamma-1}} \left(\gamma\left(\frac{3}{4\pi}m_n\right)^{\frac 13}\frac{4\pi\epsilon_0k_B }{ e^2}\right)^{\frac{3\Gamma-3}{2}} A^{\frac{\Gamma-1}{2}} Z^{3-3\Gamma} T^{\frac{3\Gamma-3}{2}}.
\end{equation}
Consequently, when the ocean has an elastic crust, equation \ref{eq:estimatesigmacrust} approximates the ocean mode frequency.}

{These expressions show the functional dependence of the mode on parameters of the model. Such modes have been shown to exist in non-homogenous atmospheres as well \citep{10.2307/96596}. This analysis demonstrates the capacity of oceans to sustain modes with lower frequencies than the neutron star $f$-mode \citep[eg]{1988ApJ...325..725M, 2021MNRAS.504.1273P} regardless of ocean stratification.}

\subsubsection{Tidal Resonance Estimates}
Since we are interested in tidal resonances during compact binary inspirals,  we estimate the time before compact binary merger of an ocean tidal resonance. Tidal resonances should occur when
\begin{equation}
   \label{eq:resonancecondition}
   \Dot{\Phi} = \frac{\omega}{m},
\end{equation}
where $\Dot{\Phi}$ is the orbital frequency of the compact binary and $m$ is the spherical harmonic index.
For circular binaries, we have 
\begin{equation}
    \label{eq:orbitalfrequencyD}
    \dot{\Phi}=\sqrt{\frac{G(M+M_*)}{D^3}},
\end{equation}
where $M_*$ is the mass of the companion object and $D$ is orbital separation. The time to merger for a given orbital separation $D$ is \citep{1964PhRv..136.1224P}
\begin{equation}
   \label{eq:mergertime}
    t_m=\frac{D^4}{4 \beta},
\end{equation}
where $\beta$ is 
\begin{equation}
    \beta=\frac{64}{5} \frac{G^3M M_* (M+M_*)}{c^5},
\end{equation}
where $c$ is the speed of light.
Combining equations \ref{eq:estimatesigma}, \ref{eq:resonancecondition}, \ref{eq:orbitalfrequencyD}, and \ref{eq:mergertime} gives an expression for the {time before merger when resonance occurs (hereafter resonance time)} in the rigid crust case as 
\begin{equation}
\begin{split}
    \label{eq:mergerestimate1}
    t_r\sim\frac{1}{4 \beta}&\left (\frac{R_{\star}^2 m^2G(M+M_*)(\Gamma-1)}{l(l+1)K\Gamma} \right)^{\frac 4 3}\\&\times\left(\gamma\left(\frac{3}{4\pi}m_n\right)^{\frac 13}\frac{4\pi\epsilon_0k_B }{ e^2}\right)^{4-4\Gamma}A^{\frac{4-4\Gamma}{3}} Z^{8\Gamma-8} T^{4-4\Gamma}.
\end{split}
\end{equation}
{Combining equations \ref{eq:estimatesigmacrust}, \ref{eq:resonancecondition}, \ref{eq:orbitalfrequencyD}, and \ref{eq:mergertime} gives an expression for the resonance time in the elastic crust case 
\begin{equation}
\begin{split}
    \label{eq:mergerestimate2}
    t_r\sim\frac{1}{4 \beta}&\left (\left(\frac{p_o}{\breve{\mu}}\right)\frac{R_{\star}^2 m^2G(M+M_*)(\Gamma-1)}{l(l+1)K\Gamma} \right)^{\frac 4 3}\\&\times\left(\gamma\left(\frac{3}{4\pi}m_n\right)^{\frac 13}\frac{4\pi\epsilon_0k_B }{ e^2}\right)^{4-4\Gamma}A^{\frac{4-4\Gamma}{3}} Z^{8\Gamma-8} T^{4-4\Gamma}.
\end{split}
\end{equation}}
These analytical estimates for the mode frequency and resonance time will allow for parameter extraction, should tidal resonances from these modes be observed.

\subsection{Ocean Mode Results}
\label{subsec:Omodes}
We now discuss the numerical values we choose for model parameters and present the computed mode results.
\subsubsection{Neutron Star Model Parameters}
\label{subsubsec:parameters}
Our neutron star model has a central density $\rho_c=10^{15}$ g cm$^{-3}$. We choose $\Gamma=2$ as was done by \cite{2021MNRAS.504.1273P}.
The value of $K$ that we use is $K=6.67\times10^{4}$ cm$^5$ g$^{-1}$ s$^{-2}$. These choices yield a neutron star that has radius $R_{\star}=12.5$ km and a mass $M=1.25$ M$_\odot$. This is just smaller than the peak mass of the galactic neutron star population 1.39 M$_\odot$ \citep{2016arXiv160501665A, 2018MNRAS.478.1377A}.
For our computations, we fix the temperature $T=10^8$ K at the crust-ocean interface, {so that $t_r< 100$ yr for all scenarios considered. A longer resonance time would be practically too long for the coincident detection of tidal resonances with compact binary mergers.} 
We note that to get temperatures as hot as $T=10^8$ K in the crust, one typically needs heating due to accretion \citep{1984ApJ...278..813F, 1990A&A...227..431H, 2003A&A...404L..33H, 2008A&A...480..459H}. For this simple study, we do not account for accretion when computing the tidal wave amplitudes or energies.

The ratio of the ocean floor depth to the neutron star radius is independent of the choice of $K$ in the equation of state.  From equation \ref{eq:oceandepth}, we find that the ocean floor depths of the three oceans are $h_{o,c}=1.14 \times10^{-4}R_{\star}$, $h_{o,o}=2.71\times10^{-5} R_{\star}$, and $h_{o,i}=8.03\times10^{-8}R_{\star}$ for carbon, oxygen, and iron, respectively.

The carbon and oxygen oceans form below the electron capture density of those elements, so the bottoms of these oceans would be a dense plasma of ions and electrons \citep{1995ApJ...449..800B}. For simplicity, we neglect the effect of the ocean having distinct layers and leave this to future work.

\subsubsection{Neutron Star Ocean Modes}
We solve equations \ref{eq:ODEs} using a four-stage Runge-Kutta scheme. We use a shooting method \citep{press1986numerical} to obtain the frequencies of each mode. {The unphysical nature of our neutron star model at the surface (i.e. that both $p=0$ and $\rho=0$ at $R_\star$) causes a divergence in equation \ref{eq:ODEs}. To avoid this divergence, we must choose a coordinate $r$ just below $R_\star$ at which to impose the surface boundary condition equation \ref{eq:surfacerel}. This ensures that our neutron star ocean model is well behaved throughout the region in which we solve the hydrodynamic equations. \cite{1995ApJ...449..800B} and \cite{2005ApJ...619.1054P} each address this, with \cite{1995ApJ...449..800B} choosing to apply the boundary condition at density $\rho=10^7$ g cm$^{-3}$ and \cite{2005ApJ...619.1054P} choosing to apply the boundary conditions at column depth $10^7$ g cm$^{-2}$. In the present work, we apply the surface boundary condition at the radial coordinate corresponding to $p=0.05p_o$. Our mode frequency calculations are robust in the following sense: applying the surface boundary condition for five different cutoffs ($p=0.05p_o$, $p=0.01p_o$, $p=0.005p_o$, $p=0.001p_o$ and $p=0.0001p_o$), we find that the computed mode frequencies change by order unity. Present limitations in the theory of neutron star oceans and atmospheres prevent achieving mode frequency calculations more accurate than within an order of magnitude.}  

Because tidal forces correspond to $l\geq2$ spherical harmonics, $l=0$ and $l=1$ modes remain practically unaffected by tidal forces, so we do not solve for them. We only solve for $l=2$ modes as those are the modes most likely to be excited tidally. As previously mentioned, we assume the neutron star is {barotropic} so that $\Gamma_1=\Gamma=2$. {As such, our ocean is unstratified and cannot sustain $g$-modes in the traditional sense (i.e. where the equation of state of the perturbed fluid differs from the background equation of state)}. 

\begin{table*}
\setlength{\tabcolsep}{12pt}
\begin{tabular}{ p{6cm} c c c}
 \hline\hline
 Ocean Makeup     & Carbon & Oxygen & Iron\\
 \hline
 Z & 6& 8& 26 \\
 A & 12 & 16 & 56\\
 Melting Density at $T=10^8$ K (g cm$^{-3}$) & $1.10 \times10^{11}$ & $2.71 \times 10^{10}$ &$8.03\times10^7$ \\
 Ocean depth $h_o$ ($R_{\star}$) & $1.14 \times10^{-4}$ & $2.71\times10^{-5}$ & $8.03\times10^{-8}$\\
 Ocean depth $h_o$ when $R_{\star}=$12.5 km (cm) & 143 & 33.9 & $1.01 \times10^{-1}$ \\
Analytic Angular frequency $\omega_i$ (s$^{-1}$)  & 241 & 117 & 6.40\\
 Numerical Angular frequency $\omega$ (s$^{-1}$)  & 104.7 & 50.98 & 2.778\\
Mode frequency $f$ (Hz) & 16.66 & 8.114 & 0.4422\\
Crust-penetrating Angular frequency $\omega$ (s$^{-1}$)  & 10.47 & 5.098 & 0.2778\\
Crust-penetrating Mode frequency $f$ (Hz) & 1.666 & 0.8114 & 0.04422\\
$\Tilde{Q}_n$ & $1.2897\times10^{-4}$& $3.0604\times10^{-5}$ & $9.0901\times10^{-8}$ \\
$H_n$ (g cm$^2$) & $2.738\times10^{41}$& $6.494\times10^{40}$ & $1.929\times10^{38}$ \\
 \hline\hline
\end{tabular}
\caption{The properties of the three neutron star oceans we consider in this work. We show the atomic number and mass of each element that we consider as the dominant substance in a $T=10^8$ K neutron star crust as well as the densities at the ocean floors calculated from equation \ref{eq:odensity}, the ocean depths from equation \ref{eq:oceandepth}, {the analytic angular mode frequency from equation \ref{eq:incomprressibleo},} the numerically computed angular frequencies and mode frequencies of each ocean, {the numerically computed angular frequencies and mode frequencies scaled by $\sqrt{\frac{\breve{\mu}}{p}}=0.1$ to estimate the crust-penetrating mode frequencies}, the dimensionless overlap integrals discussed in section \ref{sec:Tides}, as well as the integral $H_n$, defined in section \ref{sec:GWs}, which is related to the quadrupole moment of the mode. The ocean depths, mode frequencies, and integrals are all specific to this choice of equation of state and neutron star core density.}
\label{table:1}
\end{table*}
For each ocean model, we find that the ocean can sustain one $l=2$ mode with a frequency below the orbital frequency at which two neutron stars merge ($\gtrsim10^3$ Hz) \citep{2019PhRvX...9a1001A}.
{As previously mentioned, the modes we find are not the surface $g$-modes found by \cite{1988ApJ...325..725M} and \cite{2021MNRAS.504.1273P}. Instead these modes are interface modes or $i$-modes associated with the crust-ocean interface {and ocean surface}. These modes resemble shallow ocean surface waves due to the fixed crust-ocean boundary {and free ocean surface} \citep{2005ApJ...619.1054P}.} We note that stratified models can produce $g$-modes with frequencies of order $\sim1$ Hz \citep{1995ApJ...449..800B}. Table \ref{table:1} shows the densities at the ocean floor, the depths, and the mode frequencies of the neutron star ocean models, as well as integrals computed later in the paper.  

The mode frequency increases with the square root of ocean depth {as predicted by equation \ref{eq:incomprressibleo}.} Carbon oceans have the highest mode frequency at {16.7 Hz}, while iron oceans have a mode frequency of {0.44 Hz}. We note that the fully computed mode frequencies are each a factor of 2 smaller than the rough estimates obtained from equation \ref{eq:estimatesigma}. {To determine the elastic crust $i$-mode frequencies, we scale our numerically computed frequencies by $\sqrt{\frac{\breve{\mu}}{p_0}}\sim 0.1$ \citep{2005ApJ...619.1054P}, and report these in table \ref{table:1} below the rigid crust mode frequencies.}
\begin{figure*}
    \centering
    \subfigure[]{\includegraphics[width=0.95\linewidth]{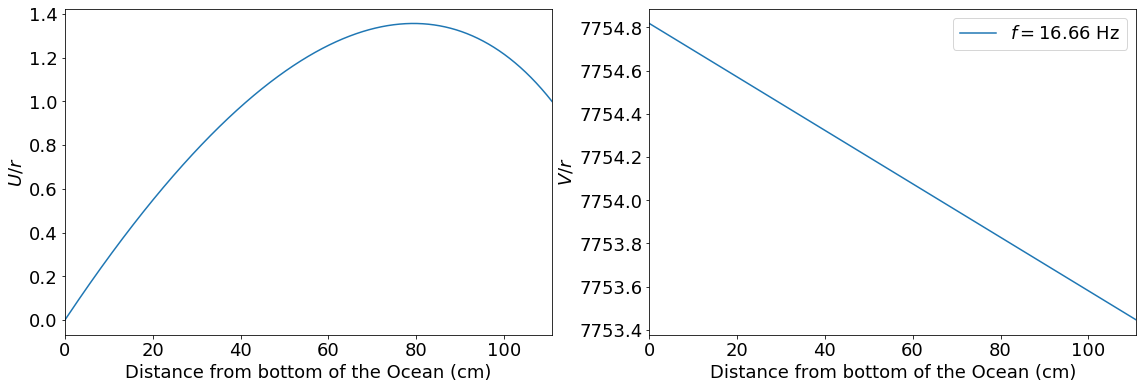}}
    \subfigure[]{\includegraphics[width=0.95\linewidth]{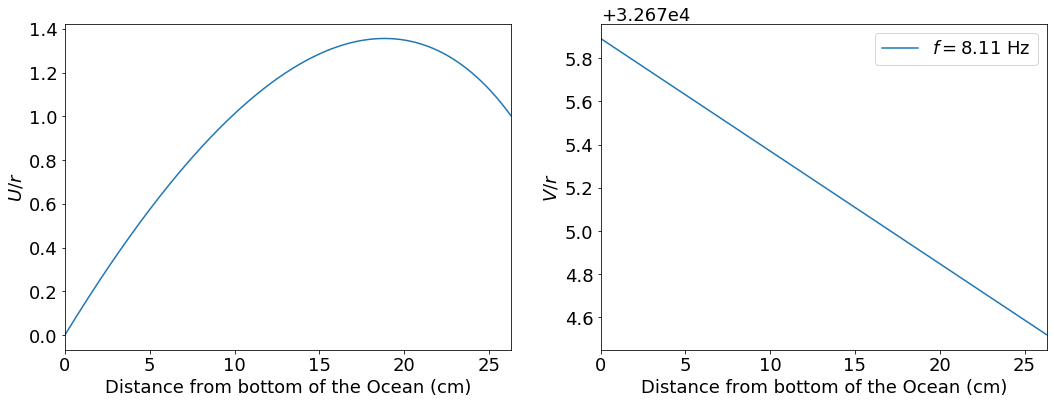}}
    \subfigure[]{\includegraphics[width=0.95\linewidth]{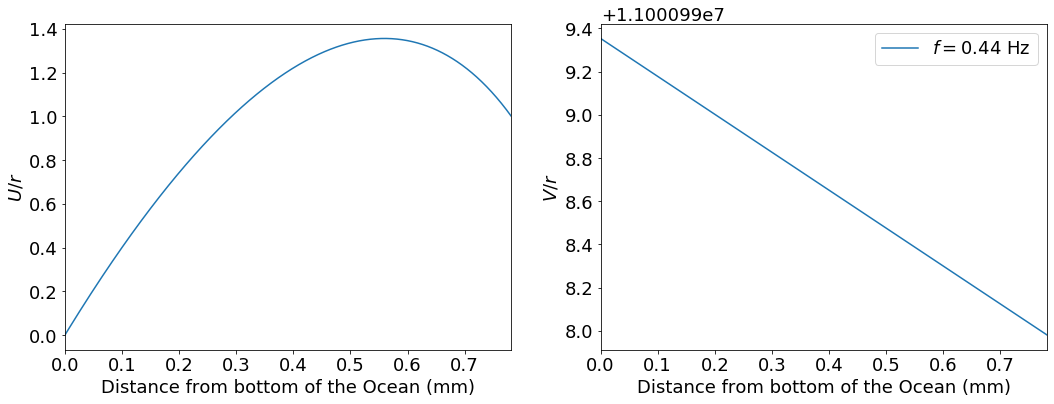}}
    \caption{The shallow ocean surface mode for the three neutron star ocean models we study in this paper: a) the carbon ocean, b) the oxygen ocean, and c) the iron ocean. The left-hand plots show the dimensionless function $\frac{U(r)}{r}$ as a function of distance from the ocean floor while the right hand plots show the dimensionless function $\frac{V(r)}{r}$ as a function of distance from the ocean floor. {Note that horizontal axis in c) is in mm because the iron ocean is only 1 mm deep.} The mode frequency of each ocean model is shown in the legend of the right hand plot.}
    \label{fig:Modes}
\end{figure*}

Figure \ref{fig:Modes} plots the radial and tangential components of the Lagrangian displacement for the three ocean modes. Each mode exhibits similar structure. The radial component $U(r)$ for each ocean has no nodes and peaks near the middle of the ocean. The tangential component $V(r)$ varies little throughout the three oceans, but well exceeds the radial component throughout. 

% -------------------------------------------
\section{Tidal Interaction}
\label{sec:Tides}
% -------------------------------------------
We now reintroduce the potential $\chi$ to equation \ref{eq:fluid}, making $\chi$ the tidal potential from a nearby companion object. The tidal potential for a companion point mass orbiting in the plane of the neutron star equator takes the form \citep{1977ApJ...213..183P, 1994MNRAS.270..611L}
\begin{equation}
\label{eq:tidal}
    \chi=-\sum_{l=2}^{\infty}\sum_{m=-l}^l \frac{G M_* r^l}{D(t)^{l+1}}W_{l m} e^{-i m \Phi(t)} Y_{l m}(\theta, \phi),  
\end{equation}
where $M_*$ is the mass of the companion, $D(t)$ is the separation between the two stars as a function of time, $\Phi(t)$ is the true anomaly, and $W_{l m}$ is the numerical coefficient \citep{1977ApJ...213..183P}
\begin{equation}
    W_{l m}=(-1)^{\frac{l+m}{2}}\frac{\left((\frac{4 \pi}{2l+1})(l-m)!(l+m)!\right)^{\frac{1}{2}}}{2^l (\frac{l-m}{2})!(\frac{l+m}{2})!},
\end{equation}
where $l+m$ must be even.

Adding the external potential $\chi$ to equation \ref{eq:fluid2} {and setting $\breve{\mu}=0$} gives
\begin{equation}
   \label{eq:fluidtide}
    \partial_t^2 \vec{\xi} + \nabla \delta \Tilde{\mu}=-\nabla \chi.
\end{equation}
Having obtained normal mode solutions to the homogenous equation $\vec{\xi}_n$, we anzats a solution to equation \ref{eq:fluidtide} \citep{1994MNRAS.270..611L}
\begin{equation}
    \vec{\xi}=\sum_n a_n(t) \vec{\xi_n},
\end{equation}
where $a(t)$ is an amplitude that scales the eigenfunction and encodes all time dependence of $\vec{\xi}$. Using the operator $\Lagr\vec{\xi}=\rho \nabla \delta \Tilde{\mu}$ defined in section \ref{subsec:ModeEOM} with $\breve{\mu}=0$, equation \ref{eq:fluidtide} becomes
\begin{equation}
    \label{eq:fluidtidewoperator}
   (\rho\partial_t^2 + \Lagr)\vec{\xi} =-\rho\nabla \chi.
\end{equation}
Substituting our ansatz for $\vec{\xi}$ gives
\begin{equation}
    \label{eq:fluidtidewoperator}
   -\rho\nabla \chi=\sum_n\rho\ddot{a}_n(t)\vec{\xi}_n + a_n(t)\Lagr\vec{\xi}_n =\sum_n(\ddot{a}_n(t) + \omega_n^2 a_n(t))\rho\vec{\xi}_n,
\end{equation}
where the last equality follows from equation \ref{eq:operator}. We use the orthogonality condition in equation \ref{eq:norm} to isolate an equation for $a_n(t)$. Applying orthogonality yields
\begin{equation}
\label{eq:amplitude1}
    \ddot{a}_n(t)+\omega_n^2 a_n(t)=-\frac{1}{A^2_n}\int\rho \vec{\xi}^*_n\cdot \nabla \chi dV.
\end{equation}
Inputting the tidal potential from equation \ref{eq:tidal}, equation \ref{eq:amplitude1} becomes
\begin{equation}
\label{eq:amplitude2}
    \ddot{a}_n(t)+\omega_n^2 a_n(t)=\frac{GM_*}{D(t)^{l+1}} W_{l m}\frac{Q_{n}}{A_n^2} e^{- i m \Phi(t)} ,
\end{equation}
where $Q_{n}$ is the overlap integral defined by \citep{1977ApJ...213..183P, 1994MNRAS.270..611L, 2012PhRvL.108a1102T,  2013ApJ...777..103T, 2020PhRvD.101h3001A, 2021MNRAS.504.1273P}
\begin{equation}
\label{eq:overlap}
    Q_{n}=\int\rho \vec{\xi}^*_n\cdot \nabla(r^l Y_{l m}(\theta, \phi)) dV=l\int\rho(U+V(l+1))r^{l+1} dr,
\end{equation}
where we have used equation \ref{eq:guess} to obtain the last equality. Note that the overlap integral is entirely a property of the mode and quantifies how strongly the mode gets excited by tidal forces. We define a normalized overlap integral, dimensionless for the $l=2$ modes, as
\begin{equation}
    \Tilde{Q}_n=\frac{Q_n}{A^2_n}.
\end{equation}
In table \ref{table:1}, we report the normalized overlap integrals for each of the three ocean modes. {We must also estimate the overlap integrals for modes which penetrate into the elastic crust. \cite{2005ApJ...619.1054P} determine that the mode energy is principally confined to the ocean, even when the mode penetrates into the crust. Furthermore, while the radial displacement has a node in the ocean with an elastic crust and not with a rigid crust, the tangential displacement of \cite{2005ApJ...619.1054P} is multiple orders of magnitude larger than the radial displacement. Because our computed rigid crust modes have this same property, the large tangential displacement in the ocean will dominate the overlap integral in both cases. Consequently, we use our computed rigid crust overlap integrals to estimate the overlap integrals of $i$-modes which penetrate into the crust.} 

Following the analysis of \cite{1994MNRAS.270..611L}, we perform a change of variables to solve equation \ref{eq:amplitude2} where
\begin{equation}
\label{eq:redefine}
    a(t)=G M_*\Tilde{Q}_{n}W_{l m}b(t)e^{- i m \Phi(t)}
\end{equation}
and $b(t)$ is the new function to solve for. In terms of $b(t)$, equation \ref{eq:amplitude2} becomes
\begin{equation}
\label{eq:bfunction}
    \ddot{b}-2 i m \dot{\Phi} \dot{b} +(\omega^2-m^2\dot{\Phi}^2-im\ddot{\Phi})b=\frac{1}{D(t)^{l+1}}.
\end{equation}
If we decompose $b$ into a real part $b^r$ and an imaginary part $b^i$, equation \ref{eq:bfunction} becomes the following two equations:
\begin{subequations}
\begin{equation}
\label{eq:breal}
    \ddot{b}^r+2 m \dot{\Phi} \dot{b}^i+m\ddot{\Phi}b^i +(\omega^2-m^2\dot{\Phi}^2)b^r=\frac{1}{D(t)^{l+1}},
\end{equation}
\begin{equation}
\label{eq:bim}
    \ddot{b}^i-2 m \dot{\Phi} \dot{b}^r-m\ddot{\Phi}b^r +(\omega^2-m^2\dot{\Phi}^2)b^i=0.
\end{equation}
\end{subequations}
Given an orbital trajectory for a companion celestial body, equations \ref{eq:breal} and \ref{eq:bim} can be solved. By plugging solutions to equations \ref{eq:breal} and \ref{eq:bim} back into equation \ref{eq:redefine}, the tidal wave amplitude $a(t)$ in the neutron star ocean can be found.
\subsection{Tidal Interaction Scenarios}
\label{subsec:tidalscenarios}
We consider three tidal interaction scenarios: a BNS inspiral in a circular orbit, a neutron star-black hole binary inspiral (NSBH) in a circular orbit, and an unbound parabolic encounter between two neutron stars (NSPE). While NSPEs are expected to be fairly rare \citep{2013ApJ...777..103T} due to the low presence of neutron stars predicted in stellar clusters \citep{2014MNRAS.440.2714B, 2018A&A...615A..91B, 2020ApJ...888L..10Y, 2022LRR....25....1M}, tidal interactions from these events remain relatively unexplored beyond \cite{2013ApJ...777..103T}, so we consider them in this work. In the following subsections, we enumerate the initial conditions and orbital parameters in each of these scenarios. 

\subsubsection{Neutron Star Binary and Neutron Star-Black Hole Binary}
The initial conditions and orbital motion of BNSs and NSBHs are largely the same when the orbital separation well exceeds the diameter of stellar mass black holes. Due to the lower mode frequencies of the three neutron star oceans, resonance will occur earlier in the inspiral than $f$-mode resonances. As such, we consider both BNSs and NSBHs at earlier times.

For an inspiraling circular binary, the time derivative of the true anomaly $\dot{\Phi}$ is just the orbital frequency 
\begin{equation}
    \dot{\Phi}=\sqrt{\frac{G(M+M_*)}{D(t)^3}},
\end{equation}
where $G$ is the gravitational constant, $M$ is the mass of the neutron star that is tidally perturbed, $M_*$ is the mass of the companion object, and $D(t)$ is the orbital separation as a function of time. The second derivative of the true anomaly is
\begin{equation}
    \ddot{\Phi}=-\frac{3}{2}\dot{\Phi}\frac{\dot{D}}{D},
\end{equation}
where $\dot{D}$ is the time derivative of $D$.
Due to the emission of GWs, the binary loses energy and $D(t)$ decreases over time. The separation as a function of time $D(t)$ for an inspiraling circular binary is given by \citep{1964PhRv..136.1224P}
\begin{equation}
    D(t)=\left(D_0^4-\frac{256}{5} \frac{G^3M M_* (M+M_*)}{c^5}t\right)^{1/4},
    \label{eq:Dt}
\end{equation}
where $D_0$ is the orbital separation at time $t=0$, and $c$ is the speed of light. We have neglected the effects of energy transfer to the neutron star ocean on the orbital motion because, as will be discussed in section \ref{sec:energetics}, the orbital energy will far exceed the energy transmitted to the ocean mode.

To numerically solve equations \ref{eq:breal} and \ref{eq:bim}, we must choose initial values for $b^r$, $b^i$, $\dot{b}^r$, and $\dot{b}^i$. We use the same initial conditions for circular binary inspirals used by \cite{1994MNRAS.270..611L} and start our integration at a time when the binary is very far from merging. These conditions are 
\begin{subequations}
\begin{equation}
   b^r(0)=\frac{1}{D^{l+1}(\omega^2-m^2\dot{\Phi}^2)} ,
\end{equation}
\begin{equation}
   \dot{b}^r(0)=\left[-(l+1)\frac{\dot{D}}{D}+\frac{2 m^2 \dot{\Phi} \ddot{\Phi}}{\omega^2-m^2\dot{\Phi}}\right]b^r(0) ,
\end{equation}
\begin{equation}
   b^i(0)=\frac{1}{(\omega^2-m^2\dot{\Phi})} (2 m \dot{\Phi} \dot{b}^r(0)+m\ddot{\Phi}b^r(0)),
\end{equation}
\begin{equation}
\dot{b}^i(0)\approx0.
\end{equation}
\end{subequations}
We compute $b$ for the $l=2$, $m=2$ cases, since $m=2$ and $m=-2$ modes will be equally excited \citep{1994MNRAS.270..611L}. The $m=0$ binary inspiral cases will be small compared to the $m=2$ resonant case. The $m=0$ case corresponds to static deformations of the neutron star, rather than the larger amplitude resonant oscillations. Resonance of the ocean mode with the tidal force is likely to occur in any isolated binary system containing a neutron star because the system's orbital frequency continuously evolves. We do not compute the $m=\pm1$ case since $Y_{2 \pm1}=0$.

\subsubsection{Neutron Star Parabolic Encounter}
We consider close encounters of neutron stars whose minimum distance of approach is a distance $s$. Since parabolic orbits correspond to those with an orbital eccentricity of $e=1$, the orbital separation as a function of radius is 
\begin{equation}
    \label{eq:parabolic}
    D(t)=\frac{2s}{1+\cos{\Phi(t)}},
\end{equation}
where $\Phi(t)$ here is the true anomaly for a parabolic orbit. Using conservation of angular momentum, we obtain a differential equation for the true anomaly as a function of time
\begin{equation}
    \label{eq:orbit}
    \dot{\Phi}(t)=\frac{1}{4}\sqrt{\frac{G(M+M_*)}{s^3}}(1+\cos{\Phi(t)})^2.
\end{equation}
We also obtain the second derivative of the true anomaly $\ddot{\Phi}$ by taking the derivative of $\dot{\Phi}$
\begin{equation}
    \ddot{\Phi}(t)=-\frac{1}{2}\sqrt{\frac{G(M+M_*)}{s^3}}(1+\cos{\Phi(t)})\sin{\Phi(t)}\dot{\Phi}(t).
\end{equation}
Solving equation \ref{eq:orbit} gives the true anomaly as a function of time for a parabolic orbit.

Again, we must choose appropriate initial conditions for $b^r$, $b^i$, $\dot{b}^r$, and $\dot{b}^i$ to solve equations \ref{eq:breal} and \ref{eq:bim}. For a parabolic orbit, the two bodies begin infinitely far away from one another with no speed. Thus, when the companion object is far away from the neutron star, we have $\dot{\Phi}(t)\approx0$ and $\ddot{\Phi}(t)\approx0$. When this is the case, $D(t)$ is approximately constant, so we obtain a first approximation to $b$ at large distances
\begin{equation}
    b\approx\frac{1}{\omega^2 D(t)^{l+1}}.
\end{equation}
The time derivative $\dot{b}$ becomes 
\begin{equation}
    \label{eq:bdotcond}
    \dot{b}\approx-(l+1)\frac{\dot{D}(t)}{D(t)}b,
\end{equation}
where $\dot{D}(t)$ is the time derivative of orbital separation. Taking the time derivative of equation \ref{eq:bdotcond} gives
\begin{equation}
    \label{eq:bddotcond}
    \ddot{b}\approx\left(-(l+1)\frac{\ddot{D}}{D}+(l+2)(l+1)\left(\frac{\dot{D}}{D}\right)^2\right)b,
\end{equation}
where $\ddot{D}$ is the second time derivative of orbital separation. We may plug equations \ref{eq:bdotcond} and \ref{eq:bddotcond} into equation \ref{eq:bfunction} and obtain an expression for $b$ containing initial conditions for both $b^r$ and $b^i$
\begin{equation}
    \label{eq:approxb}
    b\approx\frac{1}{D^{l+1}}\frac{\omega^2-m^2\dot{\Phi}^2+i(m\ddot{\Phi}-2m(l+1)\dot{\Phi}\frac{\dot{D}}{D})}{\omega^4},
\end{equation}
where we have kept only the largest terms.
Separating equation \ref{eq:approxb} into a real and an imaginary part, we get initial conditions valid when $D(t)>>s$
\begin{subequations}
\begin{equation}
   b^r(0)=\frac{1}{\omega^2D^{l+1}} ,
\end{equation}
\begin{equation}
   \dot{b}^r(0)=-(l+1)\frac{\dot{D}}{D}b^r(0) ,
\end{equation}
\begin{equation}
   b^i(0)=\frac{1}{\omega^2} (2 m \dot{\Phi} \dot{b}^r(0)+m\ddot{\Phi}b^r(0)),
\end{equation}
\begin{equation}
\dot{b}^i(0)\approx0.
\end{equation}
\end{subequations}
We solve for $b$ in the cases where $l=2$ and $m=0$. The $m=2$ NSPE tidal amplitude will be significantly weaker than the $m=0$ amplitude as resonant oscillations during NSPEs require very specific initial conditions on the neutron star trajectories, making them less likely to be found in nature. 
% ------------------------------------------
\subsection{Tidal Results}
\label{subsec:Results}
% -------------------------------------------
We report our results for the BNS, NSBH, and NSPE cases. $a(t)$ is computed by numerically solving equations equations \ref{eq:breal} and \ref{eq:bim} for $b(t)$ and substituting $b(t)$ into equation \ref{eq:redefine}. The companion mass used in the BNS and NSPE is 1.25 $M_\odot$ and the companion mass used in the NSBH is 20 $M_\odot$. We show results for the NSPE when $m=0$ and the binary inspirals when $m=2$. We report one tidal response for each possible combination of ocean and companion orbit. Table \ref{table:2} contains the main quantitative results of this paper.

\begin{table*}
\setlength{\tabcolsep}{10pt}
\begin{tabular}{ p{3.65cm} cccccc}
 
 \hline\hline
 Ocean   & Carbon (Rigid) &Oxygen (Rigid) &Iron (Rigid)& Carbon (Elastic)&Oxygen (Elastic) &Iron (Elastic)\\
 \hline
 Energy deposited (erg) &$8.6 \times 10^{46}$& $3.8\times10^{45}$ & $1.3\times10^{40}$ & $8.6 \times 10^{44}$ & $3.8\times10^{43}$ & $1.3\times10^{38}$ \\
Time before BNS merger (min) & 5.33 & 35.33 & $8.3\times10^4$ & $2.5 \times 10^3$ & $1.6 \times10^4$ & $3.9 \times10^7$ \\
 Energy deposited in NSBH (erg) & $3.9 \times 10^{46}$& $1.7\times10^{45}$ & $5.8\times10^{39}$ & $3.9 \times 10^{44}$ & $1.7\times10^{43}$ & $5.8\times10^{37}$ \\
Time before NSBH merger (min) & 0.67 & 4.5 & $1.1 \times 10^4$ & 310 & $2.1 \times10^3$ & $4.9 \times10^6$\\
 Energy deposited in NSPE (erg) & $4.3\times10^{46}$& $2.5\times10^{45}$ & $2.3\times10^{40}$& $4.3\times10^{44}$& $2.5\times10^{43}$ & $2.3\times10^{38}$\\
 Time before NSPE (s) & 0 & 0 & 0 & 0 & 0 & 0\\
 \hline
 
 \hline\hline
\end{tabular}
\caption{The main quantitative results of this paper for the three neutron star oceans and three tidal scenarios considered. This table includes the energy deposited into each ocean due to the tide and the time at which this energy is deposited. {The energy reported for the NSPE corresponds to $s=3.4\times10^6$ cm.}}
\label{table:2}
\end{table*}

\subsubsection{Resonant Tidal Waves in Binary Inspirals}
\label{subsubsec:BITide}
Both BNS and NSBH inspirals will resonantly excite the ocean modes of their component neutron stars. It is when resonance occurs that the tidal wave achieves its maximum amplitude. 

\begin{figure*}
    \centering
    \subfigure[]{\includegraphics[width=0.68\columnwidth]{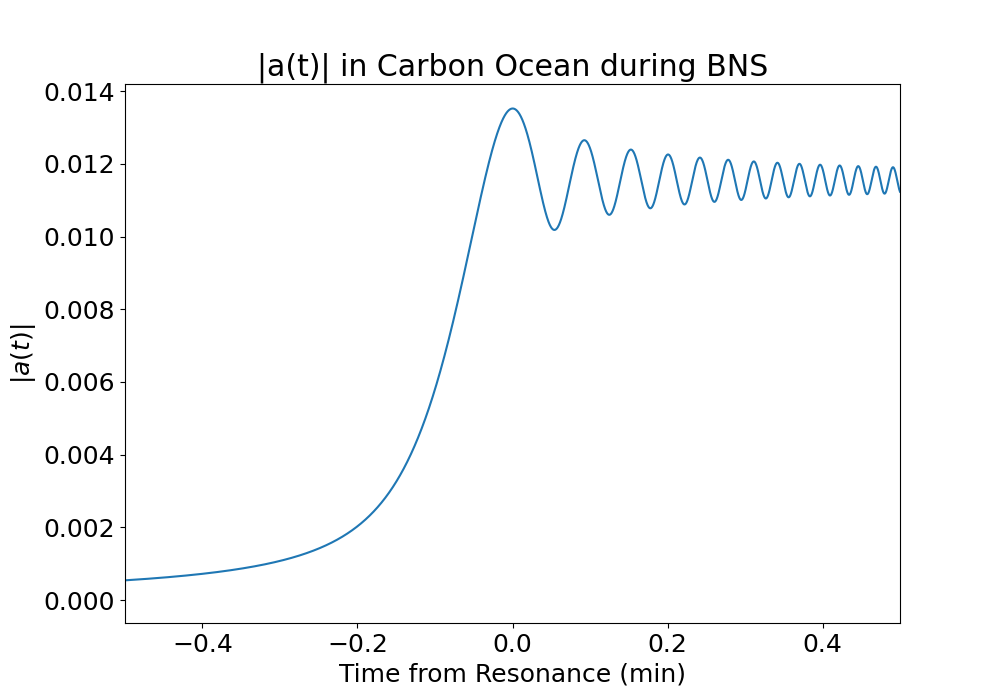}}
    \subfigure[]{\includegraphics[width=0.68\columnwidth]{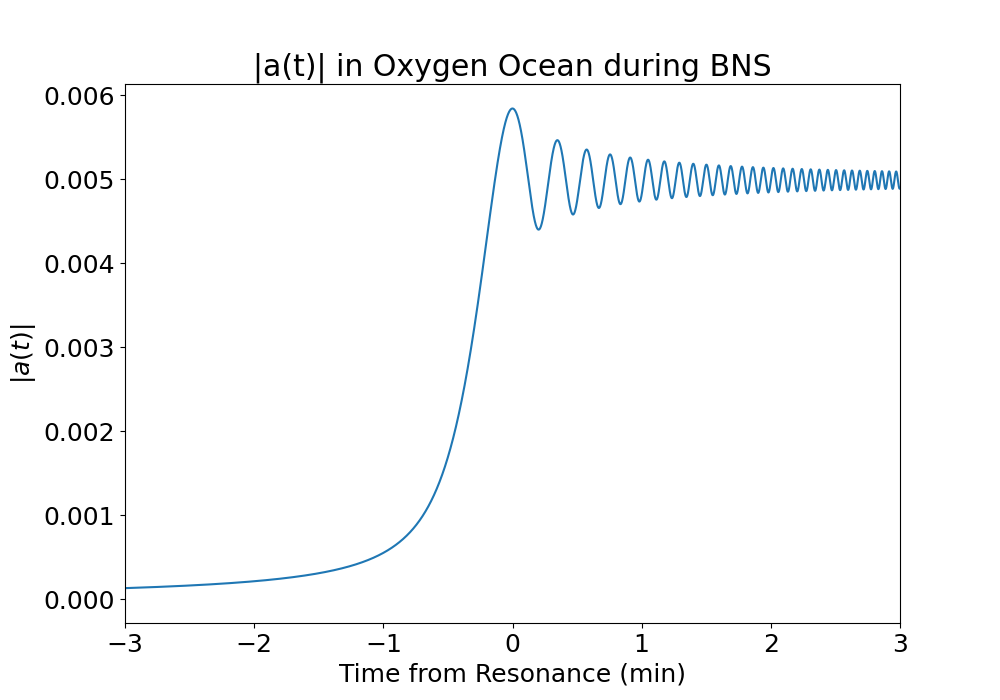}}
    \subfigure[]{\includegraphics[width=0.68\columnwidth]{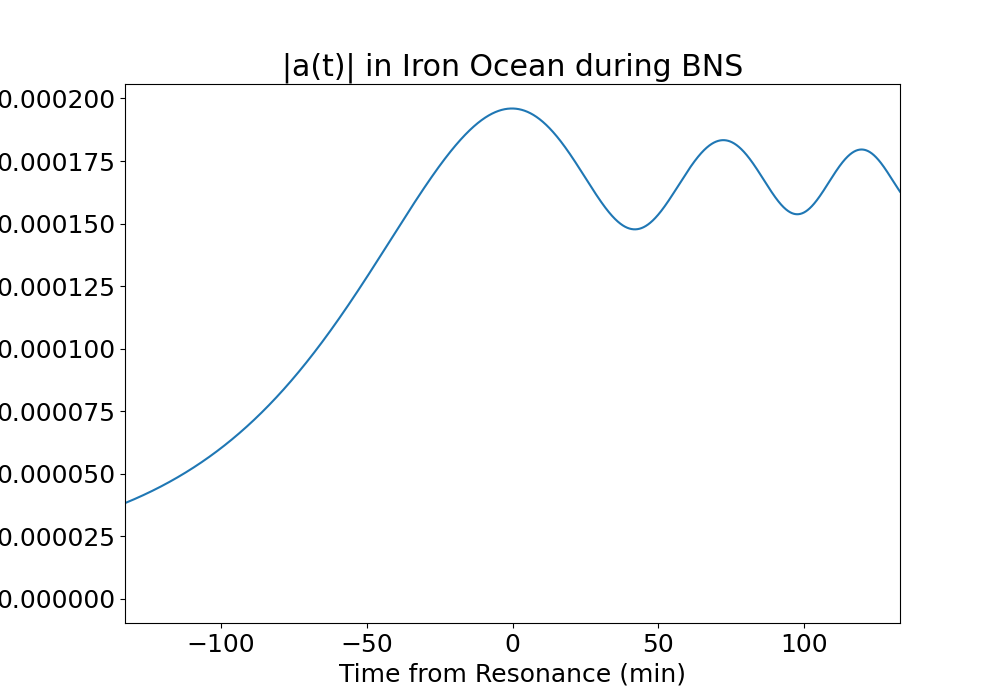}}
    \subfigure[]{\includegraphics[width=0.68\columnwidth]{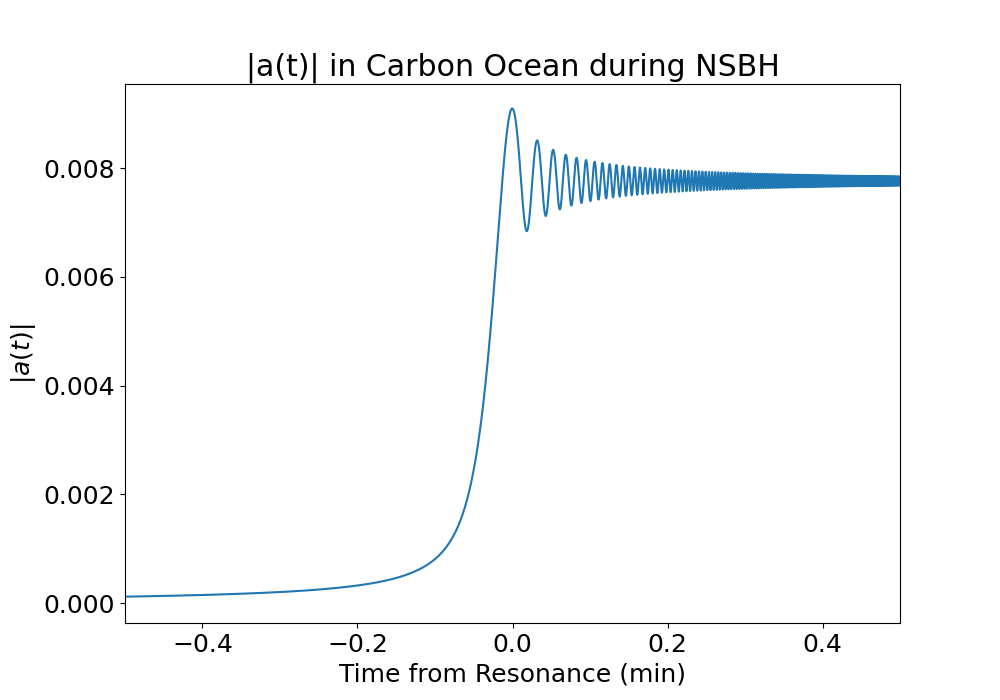}}
    \subfigure[]{\includegraphics[width=0.68\columnwidth]{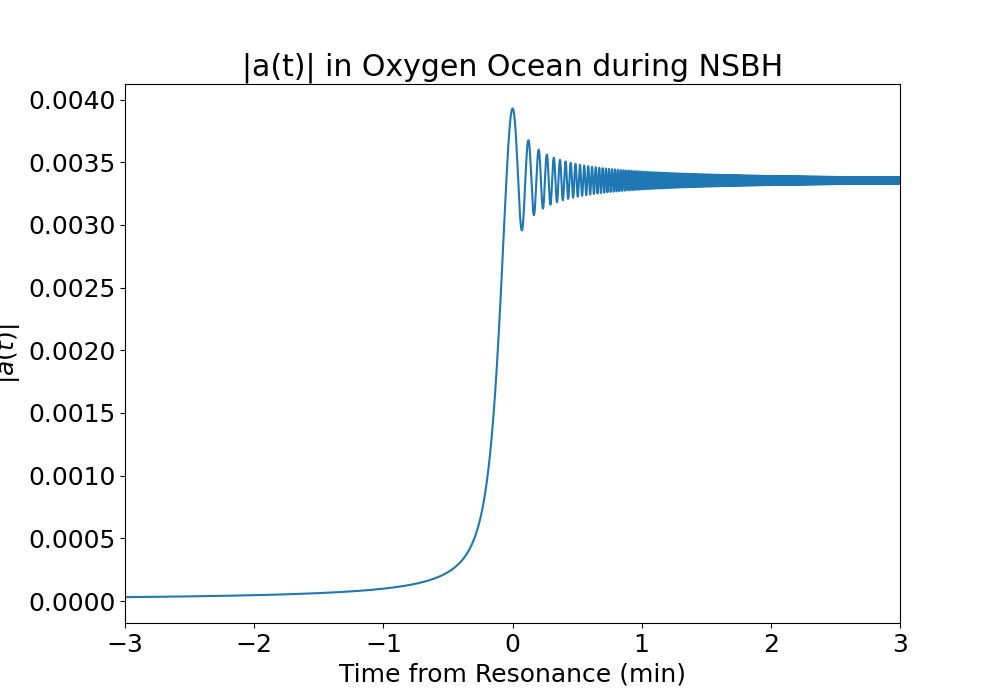}}
    \subfigure[]{\includegraphics[width=0.68\columnwidth]{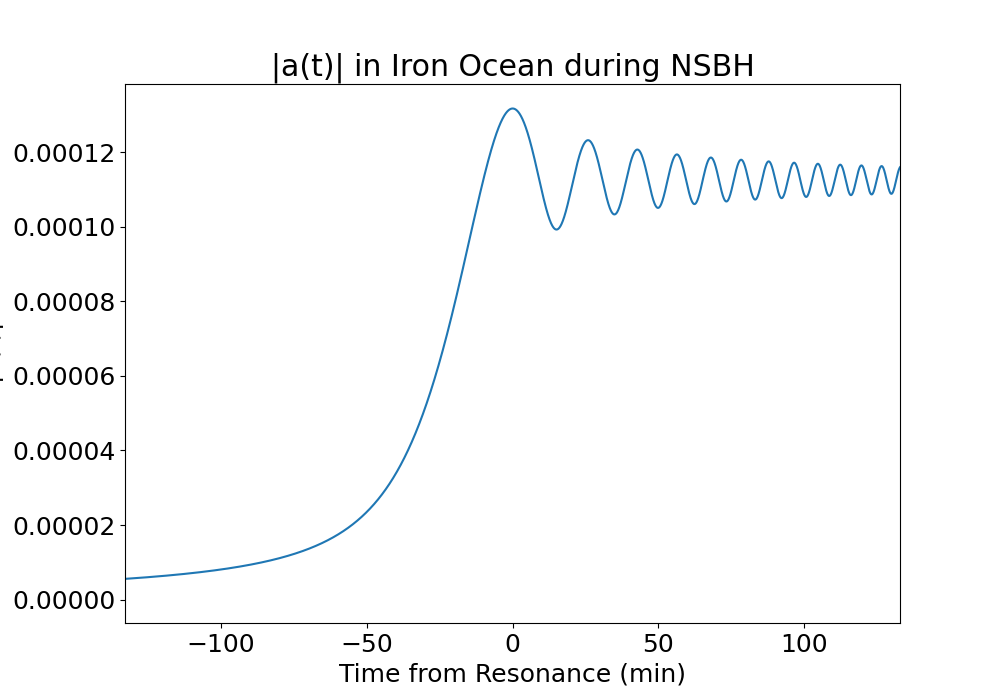}}
    
    \caption{The magnitudes of the dimensionless resonant ocean tidal wave amplitude $|a(t)|$ during compact binary inspirals as a function of time without damping. The horizontal axis shows the time from resonance in hours and the vertical axis shows $|a(t)|$. The top row shows tidal wave amplitudes during a BNS (a) in a carbon ocean, (b) in an oxygen ocean, and (c) in an iron ocean. The bottom row shows tidal wave amplitudes during an NSBH (d) in a carbon ocean, (e) in an oxygen ocean, and (f) in an iron ocean.}
    \label{fig:BItide}
\end{figure*}

In figure \ref{fig:BItide}, we show the magnitude of the tidal wave amplitudes for both BNSs and NSBHs in the times surrounding resonance. 
The general evolution of the tidal amplitudes of all three oceans is similar between both BNSs and NSBHs. In the {minutes} leading up to resonance, the amplitudes of the tidal waves increase by a full order of magnitude. We have not considered any damping mechanisms, although possible mechanisms which can decrease the tidal wave amplitudes include diffusion \citep{2021MNRAS.506L..74K, 2021PhRvD.104l3008D}, heating \citep{2016ApJ...833..261B}, and GW emission \citep{2018GReGr..50...12L}. Without damping, the tidal wave continues to pulsate with the same amplitude following the resonance time. Carbon and oxygen oceans possess tidal wave amplitudes of similar size. The overlap integrals and mode frequencies of these modes are {less than an order of magnitude different}, with carbon oceans having larger amplitudes. {In contrast, the amplitudes in the iron ocean are about a factor of 100 less than those in the oxygen ocean. These differences result from the different overlap integrals calculated for each ocean.} 

Slight differences between the BNS and NSBH case are apparent. The BNS cases generate higher amplitudes than the NSBH cases because resonance during a BNS occurs when the two bodies are roughly twice as close as during an NSBH. Additionally, the evolution of the tidal wave amplitude and frequency is noticeably slower in the BNS cases. This is a direct consequence of the slower frequency evolution in BNSs.

We determine how long before merger these resonances occur from equation \ref{eq:Dt}. We make $D_0$ the separation at resonance time, set $D(t)=0$, and solve for $t$. In BNSs {with rigid crusts}, the carbon ocean reaches resonance with the tidal force {$\sim5$ minutes} before merger, the oxygen ocean reaches resonance {$\sim40$ minutes} before merger, and the iron ocean reaches resonance {$\sim60$ days} before merger. {When scaling these results for elastic crusts, the carbon ocean reaches resonance $\sim40$ hours before merger, the oxygen ocean reaches resonance $\sim10$ days before merger, and the iron ocean reaches resonance $\sim70$ years before merger.} In NSBHs {with rigid crusts}, the carbon ocean reaches resonance {$\sim1$ minute} before merger, the oxygen ocean reaches resonance {$\sim5$ minutes} before merger, and the iron ocean reaches resonance {$\sim7$ days} before merger. {Scaling these results for elastic crusts gives resonance times $\sim 5$ hours before merger in carbon oceans, $\sim30$ hours before merger in oxygen oceans, and $\sim10$ years before merger in iron oceans.} Thus any emission from the tidally resonant oceans would well precede corresponding compact binary mergers. 

\subsubsection{Tidal Waves Excited by Parabolic Encounters}
\begin{figure*}
    \centering
    \subfigure[]{\includegraphics[width=0.68\columnwidth]{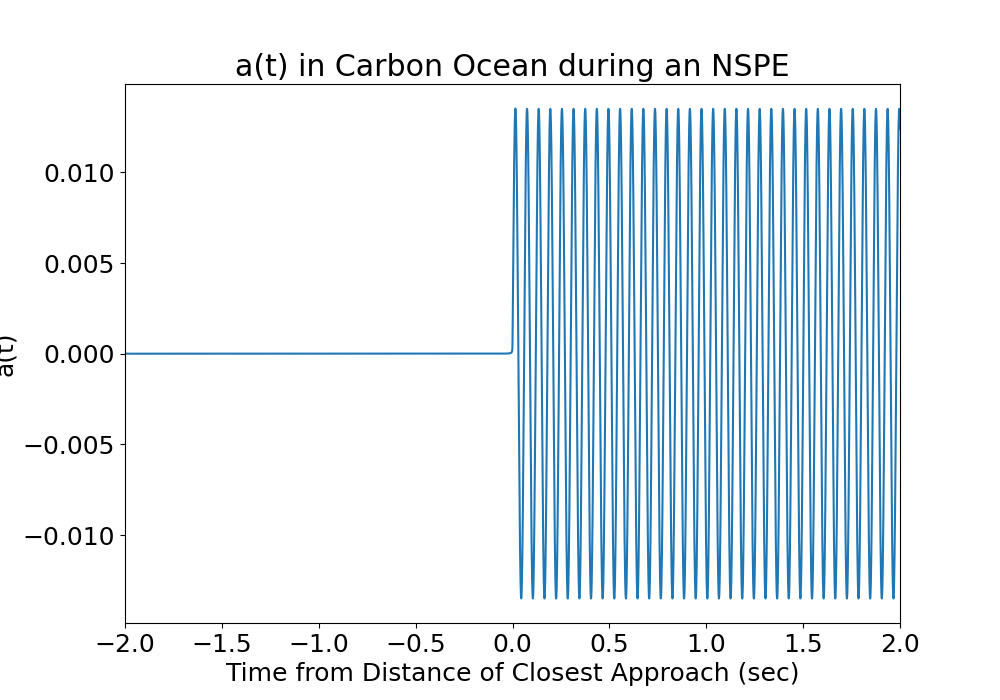}}
    \subfigure[]{\includegraphics[width=0.68\columnwidth]{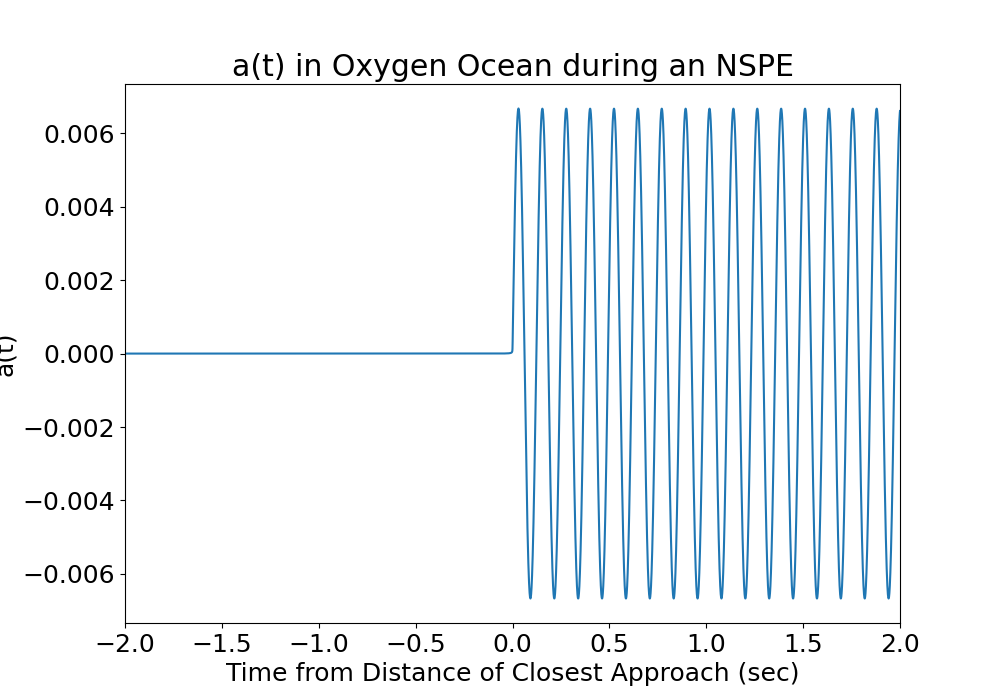}}
    \subfigure[]{\includegraphics[width=0.68\columnwidth]{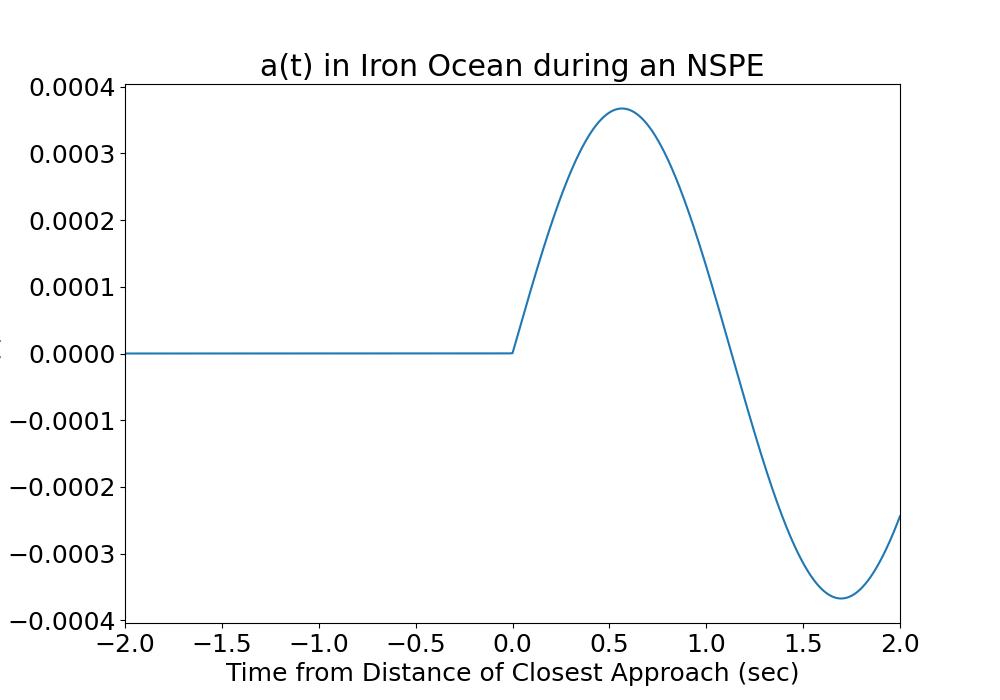}}

    \caption{The real part of the dimensionless ocean tidal wave amplitude $a(t)$ during an NSPE as a function of time without damping. The horizontal axis shows the time from resonance in seconds and the vertical axis shows $a(t)$. The sharp increase in the tidal wave height at time $t=0$ is due to the impulse from the neutron stars reaching their smallest orbital separation. (a) shows tidal wave amplitudes in a carbon ocean, (b) shows the tidal wave amplitude in an oxygen ocean, and (c) shows the tidal wave amplitude in an iron ocean.}
    \label{fig:PEtide}
\end{figure*}

NSPEs will excite tidal waves in neutron star oceans at periastron. When this occurs, the tidal force of the companion star provides an impulse to the ocean, causing it to pulsate. In this paper, we quote results when the closest distance of approach is {$s=3.4 \times 10^6$ cm}. This is the distance of closest approach where the carbon ocean tidal wave amplitude $a(t)$ in an NSPE is approximately equal to that of a BNS. For different values of $s$, the amplitudes of the excited tidal waves will scale our results by a factor of {$(s/3.4\times10^6 \text{ cm})^{-3}$}. Figure \ref{fig:PEtide} shows the tidal wave amplitudes during an NSPE for the three oceans.

After the NSPE occurs, the tidal waves will oscillate with the mode frequency of the ocean mode. The amplitudes are approximately the same order for each of the three oceans we consider. As in the binary inspiral cases, the iron ocean has the smallest amplitude. The distance of closest approach in this NSPE is about an order of magnitude smaller than the resonance distance in the binary inspiral case. NSPEs require closer encounters than NSBHs and BNSs to produce sizable ocean tidal waves. 

We estimate the event rate for NSPEs within this nominal encounter distance inside a globular cluster. NSPE event rates have been computed in previous works \citep{2006ApJ...648..411K, 2013ApJ...777..103T}, but not for these very small encounter distances. We estimate the event rate of NSPEs in a globular cluster as
{\begin{equation}
    \nu _{PE}=\frac{1}{2}N n_{\star} v_0 \sigma_{PE},
\end{equation}
where $N$ is the number of neutron stars in a globular cluster, $n_{\star}$} is the number density of neutron stars in a globular cluster, $v_0=\sqrt{\frac{G(M+M_*)}{s}}$ is the relative speed of neutron stars in an NSPE at periastron, and $\sigma_{PE}$ is the cross-section of NSPEs. Note that in this expression for event rate, we use the relative velocity between neutron stars at periastron, while \cite{2006ApJ...648..411K} use the relative velocity at infinite separation $v_\infty$. We use the velocity at periastron because we are considering parabolic orbits where $v_\infty=0$. The cross section will be
\begin{equation}
    \sigma_{PE}=\pi s^2,
\end{equation}
for the encounter distance $s$. The event rate for NSPEs within a distance $s$ becomes 
\begin{equation}
    \nu_{PE}=\frac12 \frac{N^2}{\frac43\pi R^3_{GC}}\sqrt{G(M+M_*)s^3},
\end{equation}
where we have assumed the number density of neutron stars in a globular cluster is uniform such that $n_{\star}=\frac{N}{\frac43 \pi R_{GC}^3}$ with $R_{GC}$ being the radius of the globular cluster.
Using $N=500$, $R_{GC}=1$ pc as was done by \cite{2006ApJ...648..411K}, and $M=M_*=1.25$ M$_\odot$, we find {$\nu_{PE}=5.5\times 10^{-21}$ yr$^{-1}$}. Close NSPES are therefore extremely rare.

Despite the rarity of these events, their tidal waves are generated in exact coincidence with the time of the closest passage of the neutron stars, so observation of emission from such tides can exactly demarcate the time of periastron.

\subsubsection{Energetics of Ocean Tidal Waves}
\label{sec:energetics}
The energy of an oscillation mode is divided into potential and kinetic energy. The kinetic energy and potential energy are \citep{1994MNRAS.270..611L}
\begin{subequations}
  \begin{equation}
    E_{k}=\frac{1}{2} \int \rho \frac{\partial \vec{\xi}}{\partial t} \cdot \frac{\partial \vec{\xi}^*}{\partial t} dV= \frac{1}{2} |\dot{a}(t)|^2 A^2_n,
   \end{equation}
   \begin{equation}
    E_{p}=\frac{1}{2} \int \rho \omega_n^2 \vec{\xi} \cdot \vec{\xi}^* dV= \frac{1}{2} \omega_n^2 |a(t)|^2 A^2_n.
   \end{equation}
\end{subequations}
After tidal resonance in binary inspirals, the maximum kinetic and potential energies should be equal. Additionally, both the $m=2$ and $m=-2$ modes contribute to the energy equally. Therefore, the tidal interaction will deposit a total energy into each mode \citep{1994MNRAS.270..611L}
\begin{equation}
    E=\omega_n^2|a_{n,{max}}|^2 A_n^2,
    \label{eq:modeenergy}
\end{equation}
where $|a_{n,{max}}|$ is the maximum amplitude of the tidal wave. For the NSPE $m=0$ case, only one mode contributes to the deposited energy. The NSPE total energy will be half of the energy of a binary inspiral of the same amplitude \citep{1994MNRAS.270..611L}.

We compute the energy deposited into the {shallow ocean surface mode} after tidal resonance during a BNS inspiral to be {$\sim8.6 \times10^{46}$} erg in a carbon ocean, {$\sim3.8\times10^{45}$} erg in an oxygen ocean, and {$\sim1.3\times10^{40}$} erg in an iron ocean. Similarly, we compute the energy deposited into the ocean after tidal resonance during an NSBH inspiral to be {$\sim3.9 \times10^{46}$} erg in a carbon ocean, {$\sim1.7\times10^{45}$} erg in an oxygen ocean, and {$5.8\times10^{39}$} erg in an iron ocean. The orbital energy at the time of resonance is $\gtrsim 10^{50}$ erg, justifying our assumption that the orbital motion remains unaffected.

After an NSPE whose distance of closest approach is {$s=3.4\times10^6$} cm, we compute the energy deposited into the {shallow ocean surface mode} to be {$4.3\times10^{46}$} erg for carbon oceans, {$2.5\times10^{45}$} erg for oxygen oceans, and {$2.3\times10^{40}$} erg for iron oceans. For different values of $s$, these energy results will scale by {$(s/3.4\times10^6 cm)^{-6}$}.

{The mode energy has dependence $E\propto \omega_n^2 Q_n^2$. Because the mode frequency in the elastic case is the shallow ocean surface mode frequency scaled by $\sqrt{\frac{\breve{\mu}}{p}}$, the energy deposited into the crust-penetrating $i$-modes should be the energy deposited into the corresponding shallow ocean surface modes scaled by a factor of $\frac{\breve{\mu}}{p}\sim0.01$.  \citep{2005ApJ...619.1054P}. Consequently, our energy results for the elastic crust cases are the energies reported above, reduced by a factor of 100. We also report these values in table \ref{table:1}.}

\section{Discussion}
\label{sec:discussion}
We have determined that tidal waves in neutron star oceans can be generated during BNS inspirals, NSBH inspirals, and NSPEs, and quantitatively estimated their amplitudes, energies, and timing. The tidal waves in each of these systems have unique properties. In binary inspirals, the neutron star ocean mode becomes resonant with the tidal force of the companion {minutes to days} before coalescence {if the crust is rigid, and hours to years if the crust is elastic}. Conversely, the impulsive tidal force during an NSPE excites the ocean mode at the moment of closest approach. The impulse generates simple continuous oscillatory tidal waves with the frequency of the neutron star ocean mode. The implications of these results extend to multi-messenger astronomy and neutron star geophysics. 

\subsection{Ocean Tidal Waves as Compact Binary Merger Precursor Flares and Parabolic Encounter Multi-Messenger Sources}

Dynamical activity in neutron star oceans may emit neutrino and electromagnetic radiation \citep{1994ApJ...426..688R, 2004ApJ...600..939H, 2016ApJ...832...44D, 2021arXiv210206010W}. Additionally, mode oscillations having been observed during electromagnetic bursts \citep{2014ApJ...793L..38S}. Therefore, the tidal waves in neutron star oceans during a binary inspiral might correspond to multi-messenger emission. We hypothesize that tidally resonant ocean waves in neutron stars may be a new source of compact binary merger precursor emission. 

The energies deposited into the ocean modes after resonance (computed in section \ref{sec:energetics}) represent estimates of the energy available for these flares. Thus, {$\sim10^{37}-10^{46}$} erg are available to source tidally resonant ocean flares. {The energy deposited into the carbon and oxygen oceans during NSBHs and BNSs is comparable to the breaking energy of neutron star crusts, which ranges from $10^{44}-10^{46}$ erg \citep{2012PhRvL.108a1102T, 2018MNRAS.480.5511B}. Consequently, the energy imparted to the ocean may affect the neutron star crust. If the deposited energy exceeds the breaking energy, the crust may either crack or melt. {Past work on crust breaking by resonant $i$-modes has mostly focused on the crust-core $i$-mode \citep{2012PhRvL.108a1102T, 2021MNRAS.504.1273P}. Our results show that the crust-ocean $i$-mode may have the ability to break the crust from the top, leading to interesting physics within the ocean.} Additionally, while we have neglected the presence of magnetic fields, the interaction between the excited ocean and the surface magnetic field could generate electromagnetic emission. {Particularly, if the neutron star crust breaks, subsequent magnetic reconnection of the surface magnetic field may cause large electromagnetic flares \citep[e.g.]{2015MNRAS.449.2047L, 2017ARA&A..55..261K}.} Because neutron star surfaces also emit thermal neutrinos \citep[e.g.]{2004ARA&A..42..169Y}, it is possible that this emission is manifest in neutrinos. In the remainder of this paper, we will limit our discussion to accompanying electromagnetic emission.

Pre-existing mechanisms for producing compact binary merger precursor flares include interactions of neutron star magnetospheres in BNSs \citep{2021JPlPh..87a8402A}, orbital motion of a weakly magnetized companion and a highly magnetized neutron star in either BNSs or NSBHs  \citep{1996ApJ...471L..95V,  2001MNRAS.322..695H, 2011ApJ...742...90M, 2012ApJ...757L...3L, 2012ApJ...755...80P, 2021MNRAS.501.3184S}, and tidally induced cracking of a neutron star crust during high frequency mode tidal resonances in either BNSs or NSBHs \citep{2020PhRvD.101h3002S, 2020PhRvD.101j3025G, 2021MNRAS.504.1273P, 2021MNRAS.506.2985K, 2021MNRAS.tmp.2385K}. Precursor flares from previously considered channels are only expected just before a merger ($\lesssim 10$ s) \citep{1997ApJ...482..929M, 2021MNRAS.501.3184S, 2021MNRAS.504.1273P}.

In contrast to these other mechanisms, precursor flares associated with tidally resonant neutron star ocean waves could be excited {minutes to even years} before the merger. Tidally resonant ocean flares can therefore be early warning signs of compact binary mergers involving neutron stars. Notably, NSBHs should have less trouble emitting early flares since the black hole will be farther from the neutron star than in other scenarios and should not absorb all the emission. 

Early warning precursor flares can be additional messengers for studying neutron stars and compact binary systems. The time before merger will provide information about both the type of merger and the material in neutron star oceans. In fact, the delay between flare and merger can distinguish these qualities. {Simply observing a flare within 100 years of a corresponding merger significantly constrains the parameter space and provides limits on the crust temperature. Because a crust temperature of $T\sim 10^8$ K is needed to ensure all our considered scenarios have resonance times of less than 100 years, a successful flare observation could suggest a higher crust temperature and consequently provide information about surface heating and accretion during compact binary inspirals.} 

Observing these flares in practice will likely require retroactive searches for electromagnetic data coincident in sky localization with compact binary mergers observed by GW detectors. The use of space-based GW detectors such as LISA \citep{2017arXiv170200786A} may assist in identifying flares in advance of mergers, as space-based detectors will detect GWs from compact binary inspirals well before mergers at galactic distances \citep{lisa2000lisa, 2019CQGra..36j5011R}. Observations of tidally resonant ocean flares during compact binary inspirals would complement multi-messenger efforts to study these exotic systems and their oceans. 

NSPEs could generate flares as well. The ignition of the tidal wave would precisely coincide with the NSPE. As such, coincident detections of the broadband GW bursts generated by the orbital motion \citep{1977ApJ...216..610T, 1978ApJ...224...62K, 2006ApJ...648..411K, 2012PhRvD..86d4017D} and tidally induced electromagnetic flares can allow for the multi-messenger study of NSPEs and their constituent neutron stars.

\subsection{Detection of Electromagnetic Flares from Neutron Star Ocean Tidal Waves}
We posit two possible scenarios for electromagnetic flares originating from neutron star ocean tidal waves and qualitatively discuss their detection. Since the mode frequencies of the oceans studied are {$\sim1-100$} Hz, the electromagnetic radiation from neutron star ocean tides may be ultra low frequency. As of this paper's writing, detection of ultra low frequency electromagnetic radiation on geophysical scales has been considered \citep{2002JGRE..107.5006G, 2009P&SS...57.1268G, 2016EM&P..118..103K}, but no astronomical electromagnetic instrument capable of tapping frequencies below $\sim0.001$ MHz has been proposed \citep{2009arXiv0911.0991B, saks2010daris, 2013EPSC....8..279B, boonstra2016discovering, 2018EGUGA..20.3648C, 2018ExA....46..241B, prinsloo2018emi, 2016ExA....41..271R, 2020AdSpR..65..856B}. Therefore, it would be extremely difficult to detect {$\lesssim100$} Hz radiation from neutron star ocean tidal waves.

However, due to complicated micro-physics, the large amount of energy deposited into the neutron star ocean, and the potential for magnetic reconnection, we propose that neutron star ocean tides may produce high-energy electromagnetic radiation in the gamma ray or X-ray regime with spectra and time-scales similar to that of soft-gamma repeaters (SGRs) or type-I X-ray bursts, perhaps through interactions between the surface magnetic field and the ocean. The hot temperatures of neutron star surfaces make thermal X-rays a particularly compelling manifestation of this emission. Since we have considered neutron stars with $T=10^8$ K at the crust, these neutron stars may already be accreting and emitting X-rays thermally. The tidal resonance will impart additional energy into the ocean, which we suppose may increase the flux of photons on timescales comparable to the period of the computed ocean mode. We note that accretion often requires a non-compact companion to supply material. We have neglected the effects of such additional companions for simplicity. 

Taking our high-energy flare conjecture at face-value and assuming the energy deposited into the ocean from the tide is isotropically expelled as either gamma rays or X-rays, we estimate how far away a resonant neutron star ocean tidal flare can be detected by the gamma ray detector Fermi \citep{2009ApJ...697.1071A} and X-ray telescopic array NuSTAR \citep{2013ApJ...770..103H}. 

We estimate the photon flux from such a flare by assuming all energy deposited into the mode is radiated away as either X-rays or gamma rays. Taking $R$ to be the distance between a detector and the source, we approximate the photon flux at the detector as
\begin{equation}
    F_{\gamma}\approx\frac{E}{E_\gamma}\frac{\omega}{4 \pi R^2},
\end{equation}
where $E$ is the energy of the ocean tidal wave, $E_\gamma$ is the energy of a photon, and $\omega$ is the mode frequency. We have assumed that all energy is radiated on a timescale comparable to the inverse of the mode frequency $\omega$ as it is the only short timescale we have.

For Fermi, we estimate the furthest distance at which such a flare could be observed as
\begin{equation}
    R\approx \sqrt{\frac{E}{E_\gamma}\frac{\omega}{4 \pi F_t}},
\end{equation} where $F_t$ is the photon flux threshold for Fermi. The photon flux threshold of Fermi is 0.74 photons cm$^{-2}$ s$^{-1}$ in the range of 8 keV to 40 MeV \citep{2009ApJ...697.1071A}.

For short duration X-ray sources, NuSTAR's sensitivity is limited by photon statistics. The signal to noise ratio (SNR) for a short-duration flare assuming all photons are at the same energy is 
\begin{equation}
    K=\sqrt{F_\gamma T \mathcal{A}},
\end{equation}
where $T$ is the duration of the flare and $\mathcal{A}$ is the effective area of NuSTAR. Substituting our estimate for $F_\gamma$ and taking $T\approx\omega^{-1}$ gives
\begin{equation}
    K\approx\sqrt{\frac{E}{E_\gamma}\frac{\mathcal{A}}{4 \pi R^2}}.
\end{equation}
We estimate the furthest distance at which a flare can be observed by NuSTAR as
\begin{equation}
\label{eq:distanceNustar}
    R\approx \frac{1}{K} \sqrt{\frac{E}{E_\gamma}\frac{\mathcal{A}}{4 \pi}},
\end{equation}
for some SNR threshold $K$. Notice that $\omega$ has dropped out of equation \ref{eq:distanceNustar}, so this estimate is independent of the precise timescale of the flare as long as it is short-duration \citep{2013ApJ...770..103H}.
The effective area of NuSTAR is approximately 800 cm$^2$ for photons energies of $6-10$ keV and 300 cm$^2$ for photon energies of $10-30$ keV \citep{2013ApJ...770..103H}. We set a putative SNR threshold of $K=5$. 

We compute the distances for each ocean and binary inspiral case under four detection scenarios: tidally resonant ocean flare photons are 1) gamma rays with $E_\gamma=40$ MeV detected by Fermi, 2) X-rays with $E_\gamma=8$ keV detected by Fermi, 3) higher energy X-rays with $E_\gamma=20$ keV detected by NuSTAR, and 4) lower energy X-rays with $E_\gamma=8$ keV detected by NuSTAR. {Note that $R \propto E^{\frac{1}{2}}$ for both Fermi and NuSTAR detections. Since the elastic crust energy estimates are the rigid crust energy scaled by $\frac{\mu}{p_o}$, we scale the distances from their rigid crust values by a factor of $\left(\frac{\mu}{p_o}\right)^{\frac{1}{2}} \sim 0.1$ to extrapolate the distance results for elastic crust case.} We quote our results in table \ref{table:3}. 

We find that if the emission from tidally resonant ocean flares is in the gamma ray spectrum or if crusts are composed of iron, Fermi and NuSTAR will have almost no capability to detect flares of extragalactic compact binaries, the main sources of interest for ground-based GW detectors. Excitingly, however, if tidally resonant ocean flares are emitted in the X-ray spectrum in carbon or oxygen oceans, both NuSTAR and Fermi will have the ability to detect them out to distances on the orders of {$\sim 10-1000$ Mpc for rigid crusts and even $\sim 1-100$ Mpc for elastic crusts.} These distances coincide with the BNS ranges of currently operational GW detectors \citep{2021arXiv211103606T}. In fact, BNS and NSBH inspirals have been observed out to a few 100 Mpc \citep{2019PhRvX...9c1040A, 2021arXiv210801045T, 2021arXiv211103606T}. 

We estimate an optimistic event rate for these flares using the merger rates of BNSs and NSBHs from LIGO-Virgo-KAGRA's third GW catalog \citep{2021arXiv211103634T}. The 90\% credible interval for the merger rates is reported as $10-1700$ Gpc$^{-3}$ yr$^{-1}$ for BNSs and $7.8-140$ Gpc$^{-3}$ yr$^{-1}$ for NSBHs \citep{2021arXiv211103634T, 2022LRR....25....1M}. {Assuming a flare detectable out to $\sim1-100$ Mpc, we estimate the event rates for detectable tidally resonant ocean flares by multiplying spherical volumes with radii 1 Mpc and 100 Mpc by the lower and upper limits on the quoted merger rates, respectively. The event rates would be $\sim4\times10^{-8}-7$ yr$^{-1}$ for BNSs and $\sim3\times10^{-8}-0.6$ yr$^{-1}$ for NSBHs.}
Depending on the details of the crust, ocean, and flare, precursor flares associated with tidally resonant ocean waves in compact binary inspirals may be detectable.

\begin{table*}
\setlength{\tabcolsep}{6pt}
\begin{tabular}{ p{6cm} c c c c c c}
 
 \hline\hline
 Ocean     & Carbon (Rigid) & Oxygen (Rigid) & Iron (Rigid) & Carbon (Elastic) & Oxygen (Elastic) & Iron (Elastic)\\
 \hline
 BNS Gamma Ray with Fermi (Mpc) & 3.9& 0.82 & 0.0015 & 0.39 & 0.082 & 0.00015\\
BNS X-Ray with Fermi (Mpc) & 280 & 58 & 0.11 & 28 & 5.8 & 0.011\\
 BNS Higher energy X-Ray with NuSTAR (Mpc) & 520& 110 & 0.20 & 52 & 11 & 0.020\\
BNS Lower energy X-Ray with NuSTAR (Mpc)& 1300 & 280 & 0.52 & 130 & 28 & 0.052\\
 NSBH Gamma Ray with Fermi (Mpc) & 2.6 & 0.55 & 0.0010 & 0.26 & 0.055 & 0.00010 \\
NSBH X-Ray with Fermi (Mpc) & 190 & 39 & 0.0071 & 19 & 3.9 & 0.00071 \\
NSBH Higher energy X-Ray with NuSTAR (Mpc) & 350 & 74& 0.13 & 35 & 7.4 & 0.013 \\
NSBH Lower energy X-Ray with NuSTAR (Mpc)& 900 & 190 & 0.34 & 90 & 19 & 0.034 \\
 \hline\hline
\end{tabular}
\caption{The estimated distances (in Mpc) out to which flares from neutron star oceans could be detected with Fermi or NuSTAR assuming isotropic emission. For Fermi, X-ray distances are computed assuming $E_\gamma=8$ keV, while gamma ray distances are computed assuming $E_\gamma=40$ MeV. For NuSTAR, the lower energy X-ray distances are computed assuming $E_\gamma=8$ keV, while the higher energy X-ray distances are computed assuming $E_\gamma=20$ keV. {The distances for oceans with elastic crusts are computed by scaling the distances for oceans with rigid crusts by a factor of 0.1.}}
\label{table:3}
\end{table*}

\subsection{Gravitational Waves from Neutron Star Ocean Tidal Waves}
\label{sec:GWs}
The time dependent mass density perturbations of tidal pulsations in compact stars should also generate GWs \citep{1977ApJ...216..914T}. We now investigate the GWs produced by neutron star ocean tidal waves. The GW metric $h_{ij}^{TT}$ (not to be confused with the ocean depth $h_o$) can be written as a multipole expansion \citep{1977ApJ...216..914T}
\begin{equation}
    \label{eq:GWmetric}
    h_{ij}^{TT}=\frac{G}{R c^4}\sum_{l, m} B_{lm}\left(t-\frac{R}{c}\right)T_{ij}^{lm}(\theta, \phi),
\end{equation}
where $G$ is the gravitational constant, $c$ is the speed of light, $B_{lm}(t-\frac{R}{c})$ is a time-dependent amplitude evaluated at the retarded time with dimensions of the second time derivative of the mass quadrupole moment, and $T_{ij}^{lm}$ are transverse-traceless tensor spherical harmonics \citep{1977ApJ...216..914T, 1977ApJ...216..610T}. As we have done throughout this work, we restrict ourselves to the $l=2$ harmonic. For an oscillation mode that generates small perturbations in the mass density, $B_{2m}$  is \citep{1977ApJ...216..914T}
\begin{equation}
    \label{eq:GWamplitude}
    B_{2m}(t)=\frac{16 \pi}{5 \sqrt{3}} \frac{d^2}{dt^2} \int \delta \rho Y^*_{2m}r^2 dV,
\end{equation}
where $\delta \rho$ is the Eulerian perturbation to the mass density. Rearranging equation \ref{eq:density2}, we obtain an expression for the Eulerian perturbation to the mass density
\begin{equation}
    \label{eq:Euldensity}
    \delta \rho = -\nabla \cdot (\rho \vec{\xi})=-a(t) \left(U \frac{d \rho}{dr}+\rho\left(\frac{dU}{dr}+\frac{2 U}{r}-l(l+1)\frac{V}{r}\right)\right) Y_{l m}.
\end{equation}
Substituting equation \ref{eq:Euldensity} into equation \ref{eq:GWamplitude} gives
\begin{equation}
    \label{eq:GWamplitude2}
    B_{2m}(t)=\frac{16 \pi}{5 \sqrt{3}} \ddot{a}(t)H_n ,
\end{equation}
where we have defined an integral $H_n$ as
\begin{equation}
    H_n=-\int \left(U \frac{d \rho}{dr}+\rho\left(\frac{dU}{dr}+\frac{2 U}{r}-l(l+1)\frac{V}{r}\right)\right) r^4 dr, 
\end{equation}
which quantifies an oscillation mode's ability to generate GWs.
Note that the only time dependence in equation \ref{eq:GWamplitude2} arises from $\ddot{a}$. 
We obtain results for the integral $H_n$ for each of our three ocean models. These are displayed in table \ref{table:1} in units of g cm$^{2}$. Like other integrals computed, $H_n$ is largest in carbon oceans because the carbon ocean is the largest. 

We approximate the GW strain $h(t)$ from neutron star ocean tidal waves as
\begin{equation}
    h(t)\approx\frac{G}{Rc^4} \frac{16 \pi}{5 \sqrt{3}} \ddot{a}(t)H_n. 
\end{equation}
We determine at what distance $R$ there would be GW signals with amplitudes $h \sim 10^{-20}$. This is approximately the smallest amplitude detectable with current space-based GW detector technology \citep{lisa2000lisa,2019CQGra..36j5011R}. We find that GWs from none of the configurations considered will be able to escape the immediate vicinity of the neutron star. {The values we report are for the rigid crust models.} The configuration which generates the largest GWs is a carbon ocean during a BNS inspiral. The distance from the ocean at which the GWs have an amplitude of $\sim10^{-20}$ is {$\sim10$ a.u.} In contrast, the smallest GWs are generated in an iron ocean during an NSBH inspiral. In this case, the GW amplitude is $\sim10^{-20}$ only {$\sim9$ km} away. This makes detecting GWs from neutron star ocean tides virtually impossible. While these GWs will serve as a source of extremely weak damping, we find that the damping timescales are {$\gtrsim10^7$ yr} and will not impact neutron star ocean tides on {relevant timescales.} 

While ocean tidal wave GWs may be undetectable, the orbital motion of these binary systems generates sizable GWs. During the early inspirals of BNSs and NSBHs, GWs will be detectable by LISA \citep{lisa2000lisa, 2019CQGra..36j5011R}. GWs from BNS and NSBH mergers are already detected by ground-based GW detectors \citep{2017PhRvL.119p1101A,2021ApJ...915L...5A}. Consequently, joint detection of GWs with tidally resonant ocean flares remains a possibility for multi-messenger astrophysics.

% -------------------------------------------
\section{Conclusion}
\label{sec:conclusion}
% -------------------------------------------
Neutron star oceans can sustain resonant tides. Though rather small in size, the tidal waves excited in compact binary inspirals and in parabolic encounters possess large amounts of energy, ranging of $10^{37}-10^{46}$ erg, depending on the {properties of} the neutron star crust. This energy, coupled with the rotational and magnetic energy of a real neutron star, has the potential to break neutron star crusts and fuel electromagnetic flares. Such electromagnetic flares could become early warning signs of merging NSBHs and BNS systems, {preceding mergers by $\gtrsim 1$ minute if neutron star crusts are rigid and $\gtrsim 1$ hour if the neutron star crusts are elastic.} Observations of these flares could shed light on neutron star ocean and crust properties. Their timing relative to compact binary mergers, as well as their duration and oscillation periods may serve as distinct signatures of these flares.  
Nevertheless, more work is needed to understand the physical mechanisms which can release the energy for flares as well as the effects of rotation and magnetization.

We find that tidally resonant neutron star ocean flares, if in the X-ray band, may be detected at distances of {$1-1000$} Mpc with Fermi and NuSTAR in most cases, comparable to the distances of observed BNS and NSBH mergers. We find that X-ray emission could have detection rates as high as $\sim7$ yr$^{-1}$ for BNSs and $\sim0.6$ yr$^{-1}$ for NSBHs. Neutron star ocean tides are consequently a possible source of emission which can accompany observable GWs. Subsequent work may involve reviewing past NuSTAR and Fermi data for X-ray bursts in coincident angular locations of observed BNS and NSBH mergers. 

Neutron star ocean tides and oscillations may contribute to future multi-messenger observations of astrophysical compact binary mergers and neutron stars.
Future studies into ocean tidal waves on top of crustal mountains \citep{2021MNRAS.500.5570G, 2021MNRAS.507..116G} and resultant neutron star ocean tsunamis may yield interesting results.
More exotic systems including collisions between neutron stars and planets may also produce ocean activity that results in multi-messenger emission. Multi-messenger emission from neutron stars, including emission from ocean tidal waves, will provide new knowledge about the enigmatic but rich physics of neutron stars.

% -------------------------------------------
\section*{Acknowledgments}
% -------------------------------------------
The authors are grateful to Nils Andersson, Fabian Gittins, P\'eter Petreczky, Charles Hailey, and Benjamin Owen for very helpful discussions as well as reviewing the manuscript and providing constructive feedback.
The authors thank Columbia University in the City of New York and the University of Florida for their generous support. The Columbia Experimental Gravity group is grateful for the generous support of Columbia University. A.S. is grateful for the support of the Columbia College Science Research Fellows program and the Heinrich, CC Summer Research Fellowship. L.M.B.A. is grateful for the Columbia Undergraduate Scholars Program Summer Enhancement Fellowship and the Columbia Center for Career Education Summer Funding Program. G.S. is grateful for the generous support of the Columbia University Department of Mathematics Research Experience for Undergraduates program. I.L. is grateful for the generous support of the Columbia University Amgen Scholars program. I.B. acknowledges the support of the Alfred P. Sloan Foundation and NSF grants PHY-1911796 and PHY-2110060.

\section*{Data Availability}
The data underlying this article will be shared on reasonable request to the corresponding author.

% -------------------------------------------
\bibliography{Refs} 
% -------------------------------------------
\label{lastpage}
\end{document}